\documentclass[12pt]{iopart}

\usepackage[dvips]{graphicx}
\DeclareGraphicsExtensions{.eps}
\usepackage{array}

\usepackage{version}  
\usepackage{subfigure}
\usepackage{multirow}
\usepackage[colorlinks=true]{hyperref}

\usepackage{ulem}
\usepackage{pifont}

\begin{document}
\title[]{Reconstruction of the gravitational wave signal $h(t)$ during the Virgo science runs
and independent validation with a photon calibrator}


\author{T.~Accadia$^{1}$, 
F.~Acernese$^{2,3}$, 
M.~Agathos$^{4}$, 
A.~Allocca$^{5,6}$, 
P.~Astone$^{7}$, 
G.~Ballardin$^{8}$, 
F.~Barone$^{2,3}$, 
M.~Barsuglia$^{9}$, 
A.~Basti$^{5,10}$, 
Th.~S.~Bauer$^{4}$
% ...
}
\address{$^{1}$Laboratoire d'Annecy-le-Vieux de Physique des Particules (LAPP), Universit\'e de Savoie, CNRS/IN2P3, F-74941 Annecy-le-Vieux, France}
\address{$^{2}$INFN, Sezione di Napoli, Complesso Universitario di Monte S.Angelo, I-80126 Napoli, Italy}
\address{$^{3}$Universit\`a di Salerno, Fisciano, I-84084 Salerno, Italy}
\address{$^{4}$Nikhef, Science Park, 1098 XG Amsterdam, The Netherlands}
\address{$^{5}$INFN, Sezione di Pisa, I-56127 Pisa, Italy}
\address{$^{6}$Universit\`a di Siena, I-53100 Siena, Italy}
\address{$^{7}$INFN, Sezione di Roma, I-00185 Roma, Italy}
\address{$^{8}$European Gravitational Observatory (EGO), I-56021 Cascina, Pisa, Italy}
\address{$^{9}$APC, AstroParticule et Cosmologie, Universit\'e Paris Diderot, CNRS/IN2P3, CEA/Irfu, Observatoire de Paris, Sorbonne Paris Cit\'e, 10, rue Alice Domon et L\'eonie Duquet, F-75205 Paris Cedex 13, France}
\address{$^{10}$Universit\`a di Pisa, I-56127 Pisa, Italy}
% ...
\ead{rollandl@in2p3.fr}

\begin{abstract}
The Virgo detector is a kilometer-scale interferometer for gravitational wave detection located near Pisa (Italy).
About 13~months~of data were accumulated during four science runs (VSR1, VSR2, VSR3 and VSR4) between
May~2007 and September~2011, with increasing sensitivity.

In this paper, the method used to reconstruct, in the range 10~Hz--10~kHz, 
the gravitational wave strain time series $h(t)$ from the detector signals is described.
The standard consistency checks of the reconstruction are discussed and used to estimate
the systematic uncertainties of the $h(t)$ signal as a function of frequency.
Finally, an independent setup, the photon calibrator, is described and used to
validate the reconstructed $h(t)$ signal and the associated uncertainties.

The systematic uncertainties of the $h(t)$ time series are estimated to be 8\% in amplitude.
The uncertainty of the phase of $h(t)$ is 50~mrad at 10~Hz
with a frequency dependence following a delay of $8\,\mathrm{\mu s}$ at high frequency.
A bias lower than $4\,\mathrm{\mu s}$ and depending on the sky direction of the GW is also present.

\end{abstract}


\section{Introduction}
The Virgo detector~\cite{bib:VirgoDetector}, located near Pisa (Italy), is one of the most sensitive
instruments for direct detection of gravitational waves (GW's) emitted by astrophysical 
compact sources at frequencies between 10~Hz and 10~kHz.
It is a power-recycled Michelson interferometer (ITF) with 3~kilometer Fabry-Perot cavities in the arms.\\

The four Virgo science runs (VSR1 to VSR4) accumulated a total of $\sim13$~months of data between May~2007 and September~2011,
with a sensitivity improving towards its nominal one. 
The runs were performed in coincidence with the LIGO~\cite{bib:LigoDetectors} science runs~S5 and~S6.
The data of all the detectors are used together to search for GW signals.
In case of a detection, the combined use of all the data would increase the confidence of the detection 
and allow the estimation of the GW source direction and parameters.\\

As the mirrors are moving due to environmental noises and in order to achieve optimum sensitivity,
the positions of the different mirrors are controlled~\cite{bib:LockPerformancesVSR2} to keep
beam resonance in the ITF cavities and destructive interference at the ITF output port.
The controls modify the ITF response to passing GW's below a few hundreds of hertz. 
Above a few hundreds of hertz, the mirrors behave as free falling masses; 
the main effect of a passing GW would then be a frequency-dependent variation of the output power of the ITF,
characterized by the ITF optical response.\\

The main purposes of the Virgo calibration are
(i)   to characterize the ITF sensitivity to GW strain as a function of frequency, $S_h(f)$,
(ii)  to reconstruct the amplitude $h(t)$ of the GW strain from the ITF data.
It deals with the {\it longitudinal}\footnote{
The ``longitudinal'' direction is perpendicular to the mirror surface.
} differential length of the ITF arms, $\Delta L=L_x-L_y$, where $L_x$ and $L_y$ are
the lengths of the north and west arms respectively.
In the long-wavelength approximation~\footnote{
Note that the Michelson frequency dependent response computed taking into account the finite size of the detector 
has been described in~\cite{bib:OpticalResponse}.} 
(see section 2.3 in~\cite{bib:Saulson1994}), it is related to the GW strain $h$ by
\begin{eqnarray}
h &=& \frac{\Delta L}{L_0} \quad\mathrm{where\,\,}L_0=3\,\mathrm{km} \label{eqn:LongWavelengthApprox}
\end{eqnarray}
The responses of the mirror actuation to longitudinal controls therefore needs to be calibrated, 
as well as the readout electronics of the output power and the ITF optical response.
Absolute timing of $h(t)$ is also a critical parameter for multi-detector analysis, 
especially to determine the direction of the GW source in the sky.
The calibration methods and results were described in another paper~\cite{bib:VirgoCalib}.
The scope of this paper is the reconstruction of the GW strain time series $h(t)$ from
the raw data of the ITF. 

The requirements given by the data analysis are first summarized in section~\ref{lab:AnalysisRequirements}.
After a brief description of the Virgo detector (section~\ref{lab:Virgo}) 
with the main results of the sub-system calibration, 
the reconstruction method is explained in section~\ref{lab:Method}.
In sections~\ref{lab:Checks} and~\ref{lab:Errors}, different consistency checks of the reconstructed 
time series $h(t)$ are then detailed 
and the way the systematic uncertainties\footnote{
In this paper, statistical uncertainties are given as 1\,$\sigma$ values
and systematic uncertainties as 2\,$\sigma$ values.} 
are estimated is given
along with the performances obtained during the 4th Virgo science run 
(June 3rd to September 5th 2011).
The last section is dedicated to the validation of the reconstructed $h(t)$ signal with 
an independent mirror actuation method using a setup called a {\it photon calibrator}.
Some additional studies about the control noise subtraction by the reconstruction method
are described in~\ref{lab:CtrlNoiseSubtraction}.

\section{Requirements from analysis}\label{lab:AnalysisRequirements}
The reconstructed  $h(t)$ time series is used by the data analysis algorithms to search for GW signals. 
The search sensitivity should not be limited by the uncertainties in the reconstructed times series.

The reconstructed time series $h_{rec}$ should be the sum of the possible GW signal $h_{GW}$
and the noise of the measurement, $h_{noise}$.
Reconstruction errors might lead to a bad estimation of the amplitude or of the phase
of the signals.
In the frequency domain, one can write, as a function of the frequency $f$:
\begin{eqnarray}
h_{rec}(f) &=& \Bigg( 1 + \frac{\delta A}{A}(f) \Bigg) \exp^{\mathrm{j} \delta \Phi(f)}
                           \times \Big(h_{GW}(f) + h_{noise}(f) \Big)
\end{eqnarray}
where $\frac{\delta A}{A}(f)$ is the relative error of the reconstructed amplitude,
and $\delta \Phi(f)$ is the error of the reconstructed phase.
The phase error can have two contributions: a delay $t_d$ of $h_{rec}$ over $h_{GW}$
and a frequency dependent error $\delta \phi(f)$.
This leads to $\delta \Phi(f) = -2\pi f t_d + \delta \phi(f)$. 

Three main types of error can impact the GW searches: 
(i) the statistical uncertainties, decreasing when the GW signal strength increases,
(ii) the analysis intrinsic systematic errors,
and (iii) the $h(t)$ reconstruction errors.
Taking into account the low signal-to-noise ratio of the expected GW signals in the first generation interferometers
as well as the intrinsic systematic errors, the requirements for the allowed reconstruction uncertainties
are of the order of $20\%$ in amplitude, $100~\mathrm{mrad}$ in phase and a few tens of microseconds in timing,
for each GW detector used in the analysis~\cite{bib:CalibRequirements,bib:CalibReq_Lindblom}.

\section{The Virgo detector}\label{lab:Virgo}

\begin{figure}
\begin{center}
	\includegraphics[width=0.8\linewidth]{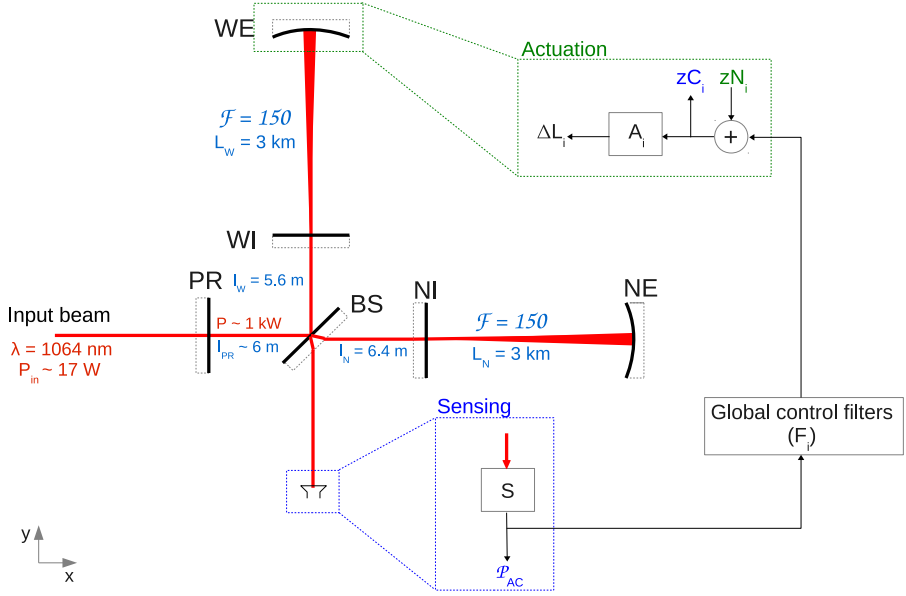}
	\caption{Optical scheme of Virgo+ and overview diagram of the main longitudinal control loop.
          For the actuation channel $i$: $A_i$ is the actuation response,
          $O_i$ is the ITF optical response to the mirror motion.
          $S$ is the transfer function of the sensing of the ITF output power $\mathcal{P}_{AC}$,
          used as error signal.
          $F_i$ is the transfer function of the global control loop.
          The actuation entries are the control signal and the calibration signal $zN_i$.
          The sum of both gives the signal $zC_i^{mir}$ (or $zC_i^{mar}$ in case of marionette actuation).
          The GW signal $h(t)$ enters the ITF as a differential motion of the two cavity end mirrors,
          filtered by the ITF optical response $O_{ITF}$.
        }
	\label{fig:OpticalScheme}
\end{center}
\end{figure}

The optical configuration of the ITF is described in figure~\ref{fig:OpticalScheme}.
A solid state laser produces the input beam with a wavelength of $\lambda=1064\,\mathrm{nm}$.
Each arm contains a 3-km long Fabry-Perot cavity which is used to increase the effective optical path.
The initial Virgo cavity finesse, $\mathcal{F}$, was~50. 
The cavity mirrors were changed during spring~2010 to obtain a finesse of about~150 in the so-called Virgo+ configuration.
The ITF arm length difference is controlled to obtain a destructive interference at the ITF output port.
The power recycling (PR) mirror sends back some light to the ITF such that the amount of light impinging
on the beam splitter (BS) is increased by a factor~40, which improves the ITF sensitivity.
The main signal of the ITF is the light power at the output port. It is called the {\it dark fringe signal}.
In practice, the ITF input laser beam is phase-modulated and the measured photodiode signal is 
demodulated to extract the light power.

The optical power of various beams and different control signals are recorded as time-series, 
digitized at 10~kHz or 20~kHz.
In order to analyze in coincidence the reconstructed GW-strain from different detectors, 
the data are time-stamped using the Global Positioning System (GPS).

\subsection{Sensing of the ITF output power}
The longitudinal control scheme adopted in Virgo is based on a standard Pound-Drever-Hall
technique~\cite{bib:LockPerformancesVSR1,bib:LockPerformancesVSR2}
and the laser beam is phase modulated. 
The main signal of the ITF is the demodulated output power, called $\mathcal{P}_{AC}$.\\

The output power of the ITF is detected using two photodiodes. Their signals then go through
analog demodulation electronics, are anti-alias filtered, digitized at $20\,\mathrm{kHz}$ and sent into a digital
process where they are summed to compute the output port channel  $\mathcal{P}_{AC}$.
In the following, the sensing transfer function from the power at the output port to the measured signal will be called $S$.
Calibration of the sensing~\cite{bib:VirgoCalib} up to $10\,\mathrm{kHz}$ results in negligible uncertainties in amplitude and phase,
except for a $4\,\mathrm{\mu s}$ uncertainty on the absolute timing with respect to the GPS time.
Note that this systematic timing uncertainty is common to all the channels recorded in Virgo.
The gain of the sensing is expected to be $1$. Possible deviations are included in the optical gain as described hereafter.

\subsection{ITF optical response: transfer function shape and optical gain}
\label{lab:OpticalResponse}
The ITF output power variations depend on the differential arm length through the so-called {\it ITF optical response} 
$G_{ITF}\times O_{ITF}(f)$ of the ITF (W/m).
$O_{ITF}(f)$ describes the frequency dependence of the transfer function while $G_{ITF}$ is the low frequency gain.\\

The Virgo detector is a Michelson-Fabry-Perot recycled interferometer. In order to increase the optical path length in the arms,
the Fabry-Perot cavities, with finesse $\mathcal{F}$ and length $L_0$, are controlled such that the beam is resonating. 
In this case, the average number of round-trips of the beam in the cavity is given by $\frac{2\mathcal{F}}{\pi}$.
Small fluctuations $\delta L$ of the cavity length induce phase fluctuations $\delta \phi_{FP}$ of the
beam reflected by the cavity, 
amplified by the number of round-trips: $\delta \phi_{FP} = \frac{4\pi}{\lambda}\frac{2\mathcal{F}}{\pi}\delta L$.

When the propagation time of the beam inside the cavity is no longer negligible with respect to the period of the length fluctuations,
the effect of the fluctuations are averaged over various round-trips. This degrades the sensitivity.
The shape of the ITF optical response has been approximated by a simple pole~\cite{bib:OpticalResponse} (with gain set to~1):
\begin{eqnarray}
O_{ITF}(f) &=& \frac{1}{1+\mathrm{j} \frac{f}{f_p}} \label{eqn:O_ITF}
\end{eqnarray}
where $f_p=\frac{\mathrm{c}}{4 \mathcal{F} L_0}$ is the cavity pole frequency
($500\,\mathrm{Hz}$ for Virgo and $167\,\mathrm{Hz}$ for Virgo+).

There is a fortuitous cancellation of the errors when combining the long-wavelength approximation
(equation~\ref{eqn:LongWavelengthApprox}) for gravitational waves and the simple pole approximation
for the interferometer response to differential length variations. 
Up to 1~kHz (respectively 10~kHz), the introduced biases\footnote{
These are the maximum bias values estimated towards the sky directions where the interferometer is the most sensitive, keeping 95\% of the total detection volume.}
are lower than 0.5\% (1\%) in amplitude.
The phase biases are almost linear with frequency and can be approximated by delays between $-4\,\mathrm{\mu s}$ and $+4\,\mathrm{\mu s}$,
depending on the sky directions: they are lower in the directions to which the detector is more sensitive to GWs,
and have thus a limited impact on the data analysis.
Note that these two approximations are also made in the LIGO $h(t)$ reconstruction.\\

The cavity pole frequency, slowly varies with time by $\pm3.5\%$. 
The main source of variation is the etalon effect in the Fabry-Perot input mirrors which have parallel flat faces.
The variations are monitored and taken into account in the $h(t)$-reconstruction process
as explained in section~\ref{lab:OptResponse}.\\

The so-called {\it optical gain}, $G_{ITF}$, in W/m, contains the gain of the ITF optical response 
and the gain of the dark fringe sensing electronics.
During VSR4, typical values were $5.7\times 10^{9}\,\mathrm{W/m}$.
It also slowly varies, in particular with the ITF alignment.
Its value is monitored from the data in the $h(t)$-reconstruction process
as explained in section~\ref{lab:OptResponse}.\\

Different optical responses $G_i\times O_i$ are defined, associated to the responses of the ITF
to variations of the positions of the different mirrors $i$.

The ITF responses to motions of the end mirrors (NE, WE in figure~\ref{fig:OpticalScheme})
and of the beam-splitter have the same shape $O_{ITF}$.
The optical gains associated with the motion of the end mirrors are expected to be equal to $G_{ITF}$,
while that associated with the beam splitter mirror is lower: $G_{BS} \sim \frac{G_{ITF}}{2\mathcal{F}/\pi}$.

The optical response to the PR mirror displacement is expected to be null.
However, due to the Schnupp asymmetry\footnote{
The lengths $l_N$ and $l_W$ are different, see figure~\ref{fig:OpticalScheme}.}, 
in the case where the north and west arms do not have the same finesse,
the optical response has the same shape as for the other mirrors, with
slight differences at low frequency (below a few tens of hertz). 
In any case, the gain of the optical response to PR displacement is low
compared to the other mirrors (below $\sim10^6\,\mathrm{W/m}$).\\

\subsection{Mirror longitudinal actuation}
For seismic isolation, all the Virgo mirrors are suspended to a complex 
seismic isolation system~\cite{bib:SeismicAttenuation,bib:LastStage,bib:PendulumTF}.
The last two stages are a double-stage system with the so-called {\it marionette}
as the first pendulum.
The mirror and its recoil mass are suspended to the marionette by pairs of thin wires. 
At both levels, electromagnetic actuators allow to move the suspended mirror along the axis of the beam.

The actuation for mirror $i$ converts a digital control signal 
(hereafter called $zC^{mir}_i$ for the mirror and $zC^{mar}_i$ for the marionette, in V)
into mirror motion $\Delta L_i$, through an electromagnetic actuator and a single or double pendulum filter.
The actuation response (hereafter called $A_i$) is defined as the product of the electronics response of the actuator 
and the mechanical response of the pendulum.
Typical gains of end mirror actuation are $0.7\,\mathrm{nm/V}$ at $100\,\mathrm{Hz}$,
with a $f^{-2}$ frequency dependence.
Once calibrated~\cite{bib:VirgoCalib}, the mirror actuation transfer function below $900\,\mathrm{Hz}$ 
is known, in modulus, to within $3\%$ and, in phase,
to within $14\,\mathrm{mrad}$ below $\sim200\,\mathrm{Hz}$ and $10\,\mathrm{\mu s}$ above.
These numbers are dominated by systematic uncertainties.

\subsection{Longitudinal control loop}\label{lab:LongLoop}

The main controlled longitudinal degree of freedom is the differential arm length (so-called {$\Delta L$}): $L_N - L_W$,
as shown in figure~\ref{fig:OpticalScheme}.
The $\Delta L$ degree of freedom is directly coupled to the dark fringe signal which senses the GW's.
The $\Delta L$ control loop used to lock the ITF on the {\it dark fringe} in science mode (standard data taking conditions)
is summarized in the figure~\ref{fig:OpticalScheme}.

The error signal is the ITF output power sensed as $\mathcal{P}_{AC}$ (W), readout with response $S$.
Filters $F_i$ (V/W) are used to define the control signals sent to the different actuation channels $A_i$ 
(m/V) in order to keep the mirrors $i$ at their nominal positions.
The NE, WE, BS and PR mirrors are controlled via the mirror actuators.
Additionally, the marionette actuators are used for the NE and WE mirrors.
The ITF output power depends on the mirror position variations through the optical response $G_{i}O_{i}$ of the ITF (W/m).\\

In 2008, it turned out that the coupling of auxiliary loop noises to the $\Delta L$ loop was not negligible.
As a consequence, the noise of the auxiliary loops would limit, below $\sim100\,\mathrm{Hz}$, 
the Virgo sensitivity when characterized directly in the frequency-domain.
Noise subtraction techniques were implemented in Virgo for all three auxiliary degrees of freedom
such that the residual motion of the auxiliary degrees of freedom does not contribute 
to limiting the detector frequency-domain sensitivity~\cite{bib:ControlNoiseSubtraction}.

\subsection{Calibration lines} \label{lab:CalibLines}
As shown in figure~\ref{fig:OpticalScheme}, a calibration signal $zN_i$ can be added to the control signal at the input of the actuation.
In Science Mode, sine wave signals are permanently sent to the different controlled mirrors: they are called {\it calibration lines}.
As explained in the following sections, they are used (i) to monitor the cavity finesse and the optical gains for the different mirrors
and (ii) to monitor the quality of the $h(t)$-reconstruction process.

The frequencies of the calibration signals that were injected during VSR4 are summarized in table~\ref{tab:CalibLines}. 
The frequency of the lines used to monitor the cavity finesse and optical gains ($\sim350\,\mathrm{Hz}$) is chosen to be at a location 
where the phase of the optical response significantly varies with the finesse 
and where the signal-to-noise ratio of the line is at least~$\sim100$ in order to have low statistical variations of the estimations.
The frequencies of the two other sets of lines used to monitor the reconstructed $ht(t)$ are chosen 
(i) in a range where the controls are applied to both mirrors and marionettas, avoiding the frequency of some interesting pulsars,
and (ii) in an intermediate range where the controls are mainly applied to the mirrors.

\begin{table}[hbp]
\begin{center}
\caption{Frequencies (Hz) and typical signal-to-noise ratio (SNR) of the calibration lines injected during VSR4.
The SNR have been estimated using FFTs of 10~s.
The SNR of the lines injected on PR was variable, depending on the finesse asymmetry in particular; 
typical values are given here.}
\begin{tabular}{|c|c|c|c|c|c|}
\hline
         &  &  \multicolumn{3}{c|}{~~~Mirror excitations~~~} & Marionette excitations \\
\hline
NE       & Freq. & 13.8 Hz  &  91.0 Hz  & 351.0 Hz            & 13.6 Hz \\
         & SNR   & 30       &  60       & 320                 & 30   \\
\hline
WE       & Freq. & 13.2 Hz  &  91.5 Hz  & 351.5 Hz            & 13.4 Hz \\
         & SNR   & 30       &  60       & 320                 & 30   \\
\hline
BS       & Freq. & 14.0 Hz  &  92.0 Hz  & 352.0 Hz            & --   \\
         & SNR   &  4       &  15       & 80                  & --   \\
\hline
PR       & Freq. & 13.0 Hz  &  92.5 Hz  & 352.5 Hz            & --   \\
         & SNR   & $\sim 3$ &  $\sim10$ & $\sim10$            & --   \\
\hline
\end{tabular}
\label{tab:CalibLines}
\end{center}
\end{table}

\section{Reconstruction method}\label{lab:Method}

\subsection{Principle}
The mirrors of the ITF are controlled to keep the detector at its operating point:
in addition to the effect of the gravitational wave signal $h(t)$,
control displacements of the different mirrors modify the differential arm length.
As a consequence, the controlled ITF has a complex frequency-dependent response to GW's.
The filters of the longitudinal control loop could be modeled to extract directly $h(t)$
from the dark fringe signal (as done in the LIGO experiment~\cite{bib:LIGO_hrec}), 
but another method is used in Virgo and is presented in this paper.
The Virgo reconstruction method for $h(t)$ is based on the subtraction of the control contributions from the dark fringe signal, 
in order to recover the signal of a free ITF.
This method makes the $h(t)$-reconstruction independent of the ITF global control system
since the knowledge and monitoring of the $F_i$ filters are not needed. 
It also allows to suppress some injected noises such as the calibration lines 
and possible noise from the auxiliary control loops.

\subsection{Main steps}

The dark fringe signal of the ITF, $\mathcal{P}_{AC}(t)$, is sensing the effective differential
arm length variations which come partly from the imposed motions of the different controlled mirrors~$i$, $\Delta L_i(t)$,
and partly from the free variations, $L_0\times h(t)$.
The variations are filtered by the frequency-dependent optical responses of the ITF $O_i(f)$.
The output power is sensed through $S(f)$.
One can write the following equation in the frequency-domain:
\begin{eqnarray}
\mathcal{P}_{AC}(f) = S(f)\times &\Bigg\{& \sum_{i} \Big[G_i O_i(f)\times \Delta L_i(f) \Big] \nonumber \\ 
                                        && \,+\,G_{ITF} O_{ITF}(f)\times L_0\times h(f)\Bigg\}
\end{eqnarray}

The mirror motions due to the control system can be computed 
from the signals sent to the mirror and marionette actuators, $zC_i(t)$,
and knowing the actuator transfer functions $A_i(f)$:
\begin{eqnarray}
\mathcal{P}_{AC}(f) = S(f)\times&\Bigg\{& \sum_{i} G_i O_i(f)\bigg[ A_i^{mir}(f) zC_i^{mir}(f) + A_i^{mar}(f) zC_i^{mar}(f) \bigg] \nonumber \\
                    &&\,+\,G_{ITF} O_{ITF}(f)\times L_0\times h(f)\Bigg\}
\end{eqnarray}
This equation can be rearranged to give $h(f)$:
\begin{eqnarray}
h(f) &=& \frac{1}{L_0 \times G_{ITF} O_{ITF}(f)} \Bigg[\frac{\mathcal{P}_{AC}(f)}{S(f)} \nonumber \\
                                      &&\quad -\,\sum_{i} G_i O_i(f)\bigg( A_i^{mir}(f) zC_i^{mir}(f) +  A_i^{mar}(f) zC_i^{mar}(f)  \bigg)  \Bigg] 
\end{eqnarray}
This equation is used to compute the $h(t)$ signal with:
\begin{itemize}
\item the time-series $\mathcal{P}_{AC}(t)$, $zC_i^{mir}(t)$ and $zC_i^{mar}(t)$  read from the raw data,
\item the mirror and marionette actuation transfer functions, $A_i^{mir}(f)$ and $A_i^{mar}(f)$, known from the actuation calibration~\cite{bib:VirgoCalib},
\item the dark fringe sensing response $S(f)$ also known from the calibration~\cite{bib:VirgoCalib},
\item the optical gains $G_i$ and responses $O_i(f)$ following equation~\ref{eqn:O_ITF}, 
with finesse and gain extracted from the data as described later in this paper.
\end{itemize}

\begin{figure}
\begin{center}
	\includegraphics[width=0.65\linewidth]{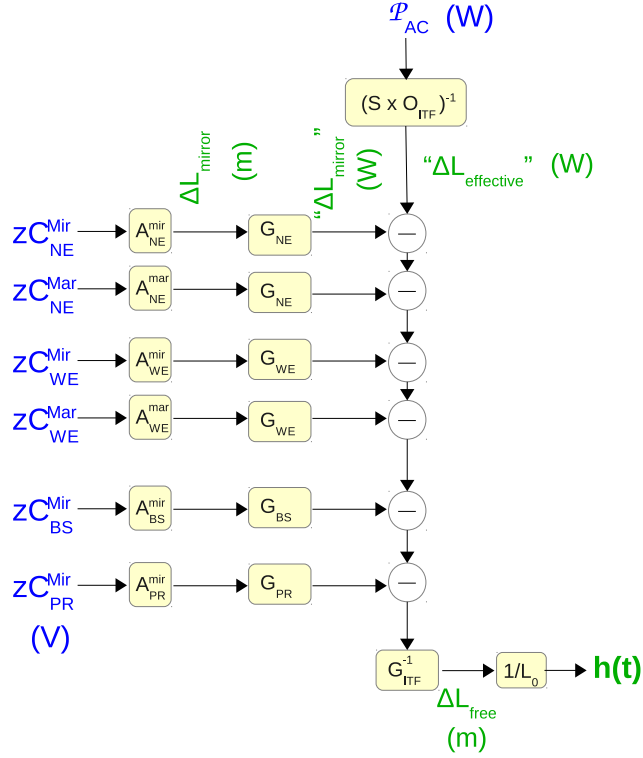}
	\caption{Principle of the $h(t)$ reconstruction.
          Blue channels are the input time-series.
	}
	\label{fig:HrecPrinciple}
\end{center}
\end{figure}

Following this equation, the principle of the $h(t)$ reconstruction algorithm is described in figure~\ref{fig:HrecPrinciple}.
The filtering can be processed in either the time-domain or the frequency-domain. 
The frequency-domain was chosen since it makes possible the correction of cavity pole and anti aliasing filters basically up to the Nyquist frequency. It also simplifies the rejection of the low-frequency band (below 10~Hz) without modifying the phase in the reconstruction band and is used to extract the optical gain on the same dataset.\\
The main steps are:
\begin{enumerate}
\item all the data are converted to the frequency-domain: Fast Fourier Transforms (FFT) are applied on all the input time-series 
with Hann window, in particular to minimize the leakage of the calibration lines which are very close.
The following steps are then performed on complex data. The FFTs are 20~s long with an overlap of 10~s between two consecutive FFTs.
\item the large low-frequency components of the data are filtered-out: a high-pass filter at 9.5~Hz is applied on all the input channels, which is a square window in the frequency-domain.
\item the inverse sensing electronics response $S^{-1}(f)$~\cite{bib:VirgoCalib} is applied on the dark fringe channel $\mathcal{P}_{AC}(f)$:
it includes the effect of the anti-alias filters and the delay to the GPS time used as reference.
\item the power variation $\Delta L_{eff}^{W}$, equivalent to the effective differential arm motion in absolute GPS time, is computed in meters:
the inverse ITF optical response $O_{ITF}^{-1}(f)$ is applied on the channel $\mathcal{P}_{AC}(f)\times S^{-1}(f)$.
\item the mirror motions due to controls, $\Delta L_i(f)$, at an absolute GPS time, are computed:
the calibrated actuation responses $A_i^{mir}$ and $A_i^{mar}$ are applied to the correction signals sent to the mirrors and marionettes
$zC_i^{mir}$ and $zC_i^{mar}$.
\item the mirror motions are converted to their equivalent dark fringe variations (W), $\Delta L_i^W$, applying the optical gains (W/m) $G_i$.
\item the power variations for a free ITF is reconstructed, $\Delta L_{free}^W$: the contributions from the actuators are subtracted from the dark fringe equivalent motion.
\item the differential arm motion for a free ITF $\Delta L_{free}$ is reconstructed applying the inverse ITF optical gain (m/W) on the previous signal. The ITF optical gain is computed as the mean of the NE and WE optical gains.
\item the result is divided by $L_0=3\,\mathrm{km}$ to get the strain $h(f)$.
\item $h(f)$ is converted back to the time-domain using inverse FFTs. To avoid glitches at the edges of two consecutive time domain segments due to the small but still present leakage effect of the FFT, the $h(t)$ stream is produced by combining time domain segments weighted by a window.
The window has been defined such that it ensures a proper normalization of the two summed signals and it smoothly starts and ends at $0$ to avoid glitches.
We checked that with this method no glitches were detected at the 10~seconds period of the FFT overlapping segments~\cite{bib:VirgoNote_Hrec_VSR1,bib:VirgoNote_Hrec_VSR2}.
\item the power lines are subtracted in the time domain as described in section~\ref{lab:PowerLineSubtraction}.
\end{enumerate}

\begin{figure}
\begin{center}
	\includegraphics[width=0.7\linewidth]{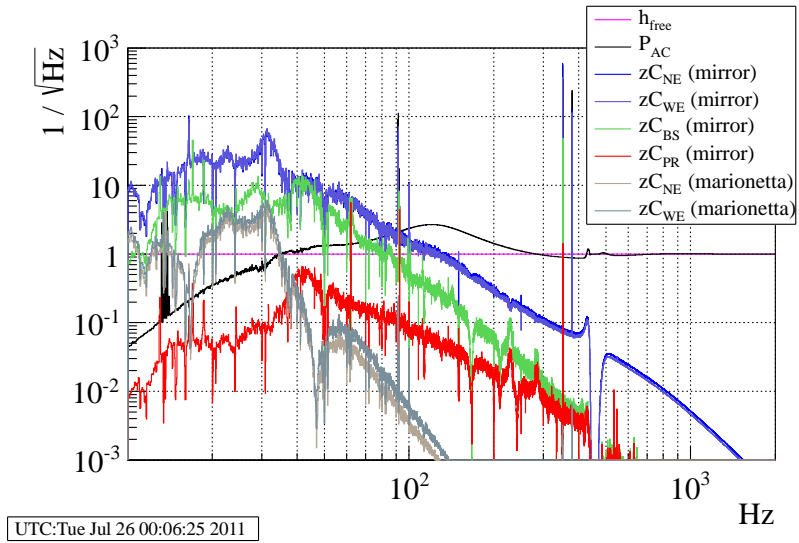}
	\caption{Spectrum of the different input time-series involved in the reconstruction, 
          normalized to the spectrum of the $h(t)$ time-series, measured during VSR4.
        }
	\label{fig:ChannelContributions}
\end{center}
\end{figure}

The relative contributions of the different input time-series to the $h(t)$ time-series are
shown in figure~\ref{fig:ChannelContributions}. 
As expected, the dark fringe signal $\mathcal{P}_{AC}$ dominates at high frequency, where the ITF is not controlled. 
In the controlled frequency band, up to a few hundreds of hertz, the main contributions come from the control signals of
the beam splitter and of the two end mirrors.
The control signals applied on the marionettes of the end mirrors contribute mainly below $50\,\mathrm{Hz}$.
The contribution from the control signals applied on the PR mirror is lower than 10\%.

\subsection{Optical responses}\label{lab:OptResponse}
The shapes of the optical responses to motions of NE, WE, BS and PR mirrors as well as the optical gains
are estimated altogether from the calibration lines.

The optical gains $G_i$ are the conversion factors from the mirror motions to the dark fringe signal 
corrected for the sensing and optical response shape.
A line used to excite a mirror will generate a line in the dark fringe signal.
Due to the control system, it will induce a correction on the other mirrors.
Therefore, one needs to take into account this correlation when extracting the optical gains of the mirrors.

Moreover, the frequency dependence $O_i$ of the optical responses to motions of NE, WE and BS mirrors is described by a simple pole
as given in equation~\ref{eqn:O_ITF}.
The slowly varying pole frequencies 
are monitored using the phase variation of the calibration lines in $\mathcal{P}_{AC}$.
It has been checked with simulations (SIESTA~\cite{bib:Siesta}) that the low-frequency difference in shape of the PR optical response is negligible, 
in particular since the optical gain of the PR response is much lower than for the other mirrors.
$O_{PR}$ is thus assumed to have the same shape as for the end mirrors in the reconstruction.

The optical responses are thus estimated solving a set of equations written at the nearby frequencies
of the calibration lines around $350\,\mathrm{Hz}$.
Assuming that the observed dark fringe signal at frequency $f_i$ is dominated by the calibration line,
it can be written as:
\begin{eqnarray}
\mathcal{P}_{AC}(f_i) &=& S(f_i)\times\sum_{j} G_j\times O_j(f_i)\times A_j(f_i)\times zC_j(f_i) \label{eqn:OpticalGains}
\end{eqnarray}
The sum is running over all the four controlled mirrors, and possibly the two controlled marionettes.

This set of equations is solved in the frequency-domain and all quantities are complex.
Therefore, the amplitudes of the unknowns give the optical gains $G_i$
while the phases allow to extract the frequency of the pole $O_i$.
Imperfections in the models $A_i$ or $S$ will of course induce uncertainties in the estimated values.\\

\begin{figure}
\begin{center}
  \includegraphics[width=0.95\linewidth]{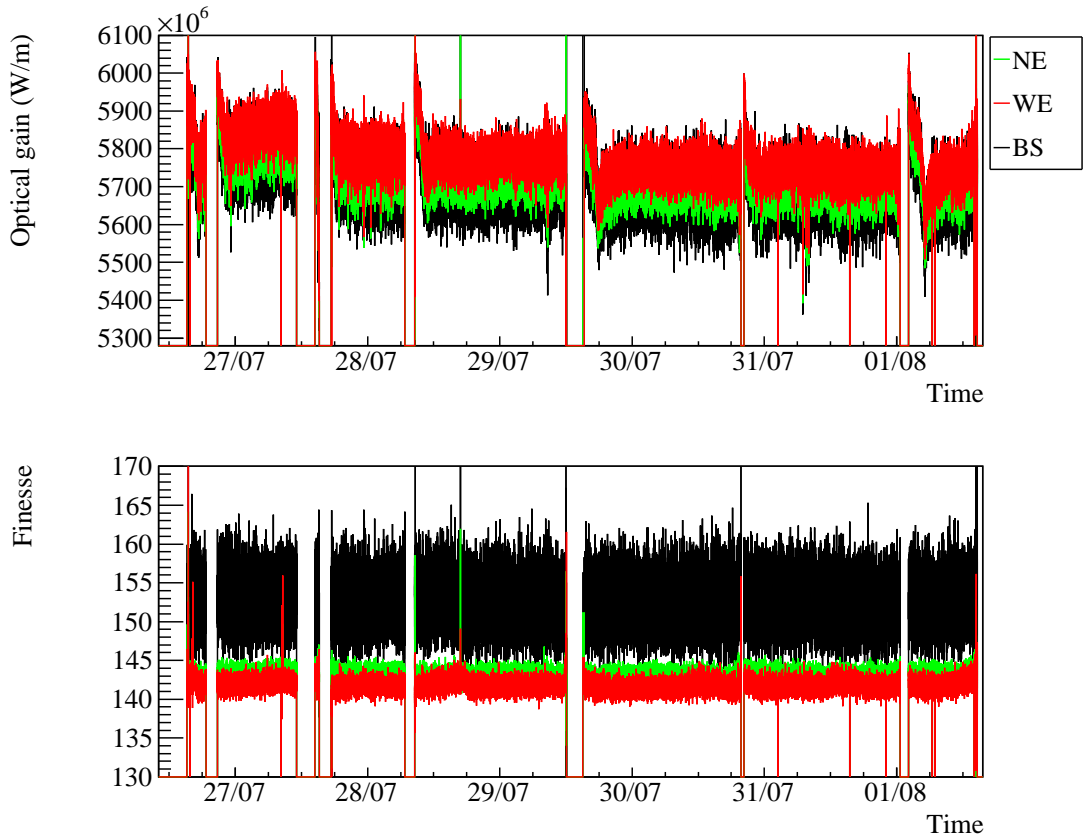}
  \caption{Optical gains and finesses estimated online by the reconstruction 
    for WE, NE and BS mirrors
    along six days during VSR4 (summer 2011).
    For better visualization, the BS optical gain have been multiplied by $2\mathcal{F}/\pi$, with $\mathcal{F}=150$.
  }
  \label{fig:OpticalGain_Finesse_VsTime}
\end{center}
\end{figure}

Such a set of equations is solved for each new FFT, i.e. every 10~s.
The extracted optical gains are then applied on the corresponding set of data.
The statistical uncertainty of the optical gain is given by the inverse signal-to-noise ratio of the lines
in the dark fringe signal. 
Their signal-to-noise ratio is of the order of~100 (see table~\ref{tab:CalibLines}), 
except for the PR line which was fluctuating over time
due to finesse asymmetry variations. However, the amplitude of the PR line was always low: which means that the 
PR coupling with the dark fringe is small, and therefore the required precision is also low.\\

The finesse (or pole frequency following equation~\ref{eqn:O_ITF}) and optical gains estimated for WE, NE and BS 
during six days of VSR4 are shown in figure~\ref{fig:OpticalGain_Finesse_VsTime}.
Different lock segments are visible.
The optical gains and finesses are estimated with statistical uncertainties of the order of 2\% for BS and 0.5\% for NE and WE.
While it is expected that the finesse measured via the BS mirror is the average of the finesse of the north and west arms,
it is not the case in the data. The difference can be explained assuming a relative error in the calibration of BS mirror actuation
with respect to the calibration of NE and WE actuation of $20\,\mathrm{mrad}$. This is well inside the systematic uncertainties
of the mirror actuation calibration estimated to $10\,\mathrm{\mu s}$ (i.e. $22\,\mathrm{mrad}$ at 350~Hz) in~\cite{bib:VirgoCalib}.
As a consequence, the systematic uncertainties on the finesse estimated in the $h(t)$ reconstruction 
are of the order of 6\%.

Finesse variations of $\pm2\%$ over the runs are due to different tuning of the etalon effect in the arm cavity input mirrors
with the thermal compensation system~\cite{bib:TCS}. The optical gain variations, also of the order of $\pm2\%$, are mainly
related to the alignment status of the ITF.

\subsection{Power line subtraction}\label{lab:PowerLineSubtraction}
The noise generated by the power supplies in Virgo is located at the mains 50~Hz frequency, and its harmonics.
A feed-forward technique is applied to reduce their contribution in the $h(t)$ signal.
The power distribution is permanently monitored in channel $P_{50Hz}$.
Different steps are performed on 1~s long time-series as described in~\cite{bib:PowerLineSubtraction}:
\begin{enumerate}
\item the frequency and phase of the 50~Hz mains are measured from $P_{50Hz}$,
\item theoretical sine waves are built using this phase and amplitude for the main signal and its first 18~harmonics.
The amplitude of the sine waves are derived from the coupling coefficient between the power line and the $h(t)$ channel
measured in the previous data segment,
\item these artificial power line signals are subtracted in the time-domain from the raw $h(t$) time-series to produce
the final ``clean'' $h(t)$ time-series,
\item the coupling between the raw $h(t)$ signal and the power line is measured to provide the coupling coefficient
for the following data segment.
\end{enumerate}

\subsection{Data quality flags and monitoring channels}
The quality of the reconstructed $h(t)$ time-series is evaluated in the reconstruction process every 10~s.
The conditions to get a good quality are:
\begin{itemize}
\item the ITF is at its standard operating point,
\item all needed time-series are available in the data for the previous, current and following 10~s frames
(this is needed by the frequency-domain filtering),
\item the signal-to-noise ratio of the NE, WE and BS calibration lines is above~3,
\item the individual finesse extracted for NE, WE and BS optical responses are all in the 100--200 range 
(for Virgo+ with nominal finesse of~150),
\item the $P_{50Hz}$ time-series used for the power-line subtraction is available in the current 10~s frame.
\end{itemize}
The results of these tests are recorded in time-series sampled at 1~Hz and stored in the data.
During the 2243~hours of the run VSR4, Virgo was in science mode 82\% of the time,
from which the overall duty cycle of the $h(t)$ reconstruction was 99.93\%.
The independent duty cycles of the $h(t)$ quality criteria are all around 99.99\%.

Some monitoring time-series are produced at 0.1~Hz and also stored in the data:
the finesses and the optical gains estimated for the PR, BS, WE and NE mirrors, 
and the averaged ITF finesse and optical gain.

\section{Consistency checks}\label{lab:Checks}
Various consistency checks are performed on the computed $h(t)$ time-series
in order to validate the sign of $h(t)$ and to estimate the systematic uncertainties
in modulus and phase. Specific data were taken every week during the Science Runs for this purpose.

\subsection{Cavity finesse} \label{lab:Check_Finesse}
The finesse of the Fabry-Perot cavities is estimated independently in the calibration process
studying the shape of the Airy peaks in dedicated data when the arm cavity mirrors are freely swinging~\cite{bib:NoteCavityFinesse}:
the finesse is estimated right after the ITF has lost its standard conditions (in order to reduce the possible
finesse variation due to thermal effects). 
Systematic errors of the order of 2\% have been estimated for this method.
The finesse estimated in the $h(t)$-reconstruction at the end of the standard condition segment
is then compared to the finesse estimated with the Airy peaks.

The two measurements are well correlated, but with a finesse offset of~$\sim5$ between the two methods during VSR3. 
Assuming the offset comes from an error in the finesse estimated during the $h(t)$ reconstruction,
its origin would be a phase error in the calibration at the frequency of the calibration line ($f_c\sim351\,\mathrm{Hz}$).
A systematic offset of $\alpha\,\mathrm{rad}$ could be interpreted as a timing mismatch
$\delta t = \frac{\alpha}{2\pi f_c}$ between the actuation and the sensing parameterizations, $A_i$ and $S$,
where $f_c$ is the frequency of the calibration line used to estimate the finesse in the reconstruction.

During VSR2, with a nominal finesse of~50, a finesse offset of 1.8 was observed, indicating a timing mismatch of $7.8\,\mathrm{\mu s}$.
Then the mirrors were changed to increase the nominal finesse to~150:
during VSR3, a finesse offset of 5 was observed, indicating a timing mismatch of $6\,\mathrm{\mu s}$.
During VSR4, the finesse could not be properly estimated from the Airy peak shapes, but the mirrors were the same as during VSR3
and it was shown that the calibration parameters had not changed between VSR3 and VSR4: as a consequence,
the same offset can be assumed for VSR4.

Such a timing error is compatible with the systematic uncertainties given on $A_i$ and $S$ by the calibration procedure.
As a consequence, a fine-tuning of the timing in the parameterizations can be done. 
Since the origin of this offset is not known, during VSR3 and VSR4, both the timing of the $S$ and $A_i$ parameterizations
were modified by $3\,\mathrm{\mu s}$ compared to the initial calibration measurements. 
The timing uncertainty estimated later on the $h(t)$ signal takes into account this fine tuning.

\subsection{Injections with out-of-loop actuators} \label{lab:HrectoHinj}
A simple way to check that the $h(t)$ signal is correctly reconstructed is
to compare it with a known $h(t)$ excitation applied to the detector.
An excitation signal $zN$ is applied to out-of-loop mirror electro-magnetic actuators.
$zN$ can be translated to a mirror displacement through the calibrated mirror actuation $A$,
or to an equivalent signal $h_{inj}$.
The transfer function from the injected displacement to the reconstructed signal,
$\frac{h_{rec}(f)}{h_{inj}(f)}$ is expected to have a flat modulus equal to~1 
and a flat phase equal to~0. 
Deviations give an estimation of the systematic uncertainties of the reconstructed $h(t)$ channel.

\begin{figure}
\begin{center}
	\includegraphics[width=0.8\linewidth]{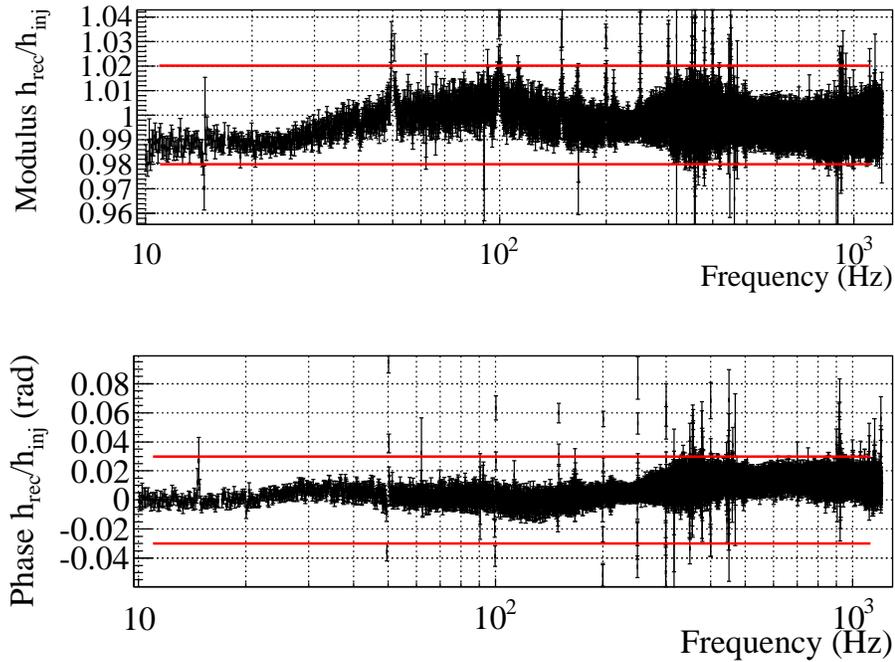}
	\caption{Average transfer function between the reconstructed $h(t)$ time series and the
          $h(t)$ signal simulated in the interferometer via electromagnetic mirror actuators.
          The average was performed on the weekly measurements of the transfer function during VSR4,
          selecting only the points with coherence higher than 95\% between both signals.
          The red lines indicate the levels of the $h(t)$ systematic uncertainties derived from the average transfer function.
        }
	\label{fig:Hrec_to_Hinj_VSR4}
\end{center}
\end{figure}

Such measurements were performed every week during the Virgo Science Runs,
$zN$ having frequency components in the range $\big[10\,\mathrm{Hz}-1\,\mathrm{kHz}\big]$.
After having checked the stability of the measurements over a run,
the weekly transfer functions were averaged to reduce the statistical uncertainties.
The results for VSR4 are shown in figure~\ref{fig:Hrec_to_Hinj_VSR4}.
Except for the frequencies of the power lines, at which a larger dispersion is observed,
the modulus is flat with a variation of no more than $\pm2\%$ around~1,
and the phase is also flat to within $\pm30\,\mathrm{mrad}$ around~0.

In the case there were a common error in the calibration of all the gains of the mirror actuator responses,
it would not be detected by this comparison of the reconstructed $h(t)$ time series with a signal
simulated through the mirror actuators: both the $h_{rec}$ and $h_{inj}$ signals would have the same 
error that would be cancelled when calculating the ratio.
In the case of a timing mismatch between the actuation and the sensing parameterizations
used in the $h(t)$-reconstruction, such transfer functions would have a non-flat shape around a few
tens of hertz, where the contributions of the control signals and of the dark fringe signal
have a similar contribution to $h(t)$. 

Note that the $h(t)$ signal is reconstructed from the dark fringe signal and the mirror control signals,
using the corresponding calibration responses without tuning, except for an additional delay between
the dark fringe and the controls.

\subsection{Noise level in the reconstructed $h(t)$ time-series} \label{lab:NoiseLevelInH}

Even if the reconstruction process produces a $h(t)$ time-series with the correct amplitude and phase,
it could still add extra-noise, in particular if the control signals are not properly cancelled-out
in the reconstruction.
On the other hand, if the online cancellation of the control signals in the detector loops is not optimal,
a proper $h(t)$-reconstruction could remove some of this control noise, as its does for the calibration lines.

In this section, studies computed on Science Run data, when the online cancellation of control signals
was efficient, are shown. 
Specific data without the online cancellation are analyzed in~\ref{lab:CtrlNoiseSubtraction}.

\subsubsection{Comparison with frequency-domain sensitivity --}
An estimation of the noise added to or subtracted from the $h(t)$-channel can be made by comparing
the $h(t)$ spectrum to the sensitivity computed in the frequency-domain as described in~\cite{bib:VirgoCalib}.
The frequency-domain sensitivity $h(f)$ is computed from the dark fringe channel $\mathcal{P}_{AC}$
to which the detector transfer function has been applied. Such sensitivity measurements were performed every week
during the science runs. Below 900~Hz, the transfer function was directly taken from the measurements, with
statistical fluctuations. At higher frequency, the transfer function cannot be measured directly: 
it was therefore extrapolated by a model fitted on the data between 900~kHz and $\sim1$~kHz
and does not contain statistical fluctuations.

\begin{figure}[bp!]
\begin{center}
\subfigure[Comparison of overlaid sensitivities.]{\label{fig:HRECsens_vs_Fsens}
	\includegraphics[width=0.8\linewidth]{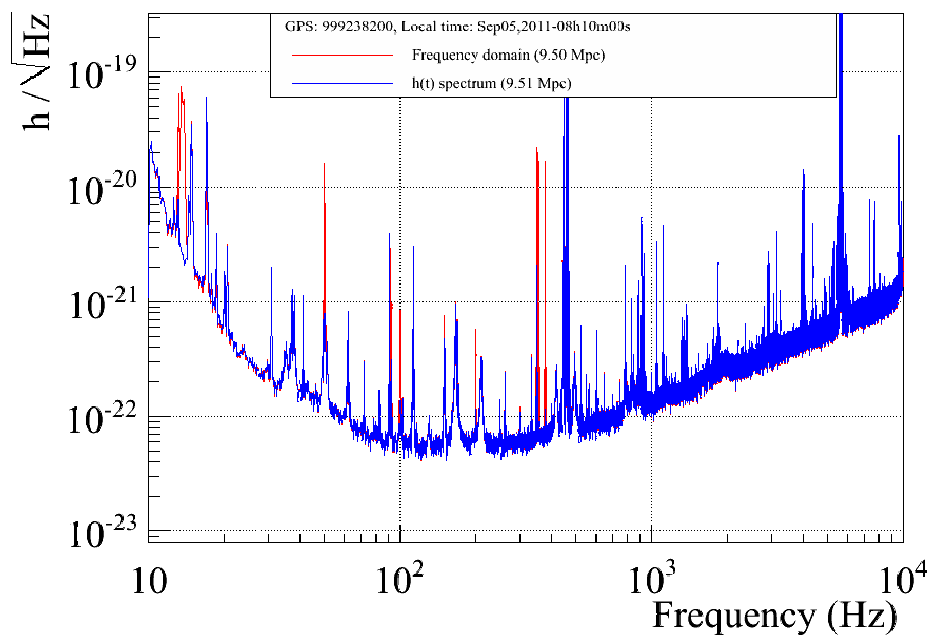} }
\subfigure[Comparison of sensitivities: ratio. variations during VSR4.]{\label{fig:HRECsens_vs_Fsens_ratio}
	\includegraphics[width=0.8\linewidth]{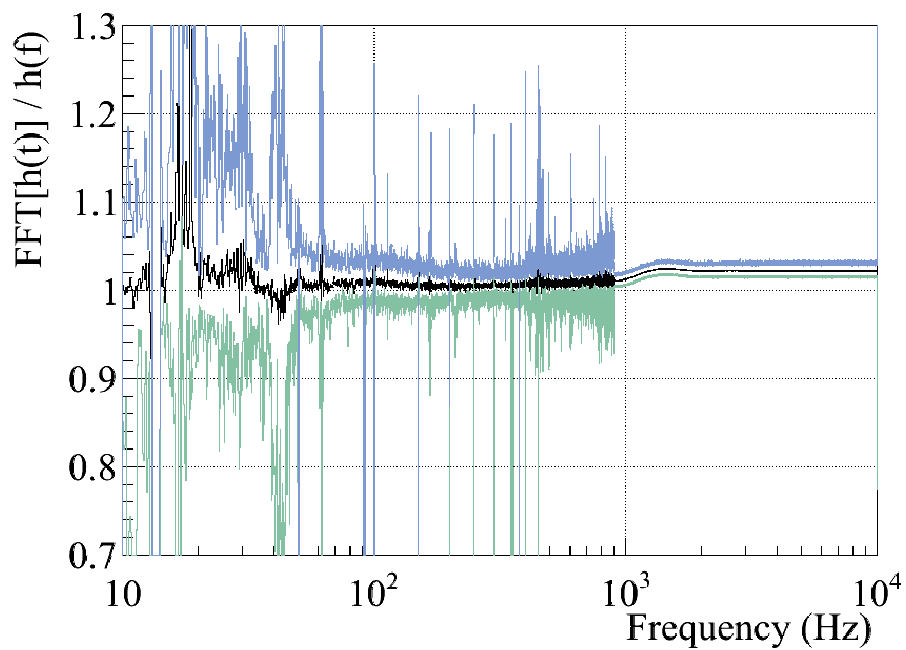} }
	\caption{Comparison of detector sensitivity curves estimated during VSR4
          from the dark fringe signal and the interferometer closed-loop transfer function
          (red curve in (a)) and as the spectrum of the reconstructed $h(t)$ time series (blue curve in (a)).
          (b): the ratio of both sensitivity estimates has been performed on a weekly basis.
          The average ratio (black), minimum value (green) and maximum value (blue) estimated over all the VSR4 measurements are shown.
        }
	\label{fig:Sensitivities}
\end{center}
\end{figure}

The two estimations of the Virgo sensitivity are compared in figure~\ref{fig:HRECsens_vs_Fsens}.
In order to compare $FFT[h(t)]$ to $h(f)$, their ratio is calculated.
Their average, minimum and maximum values are estimated over each run and shown in figure~\ref{fig:HRECsens_vs_Fsens_ratio}.
The vertical lines indicate the power lines and the calibration lines which are subtracted 
in the reconstruction process. 
No excess noise is observed in the $h(t)$ channel.
During VSR4, various techniques of noise cancellation were applied in the control loops:
therefore, the $h(f)$ sensitivity was not limited by control noise to be subtracted when calculating $h(t)$ 
and the ratio is still close to~1 at low frequency.
The increase of the ratio by $\sim2\%$ around 1~kHz comes from a systematic error in the $h(f)$ estimation
since it assumes that the contribution of the controls are completely negligible above 900~Hz while they still contribute
at the $\sim2\times1\%$ level as shown in figure~\ref{fig:ChannelContributions}.
The change in the behavior of the noise at 900~Hz comes from the way the detector transfer function 
is estimated when computing the frequency-domain sensitivity curve as explained earlier.

\subsubsection{Coherence between $h(t)$ and the control signals --}
The main control loop of the ITF described in this paper controls the differential arm length.
Other degrees of freedom of the ITF are controlled to keep it at its operating point:
the differential length of the short Michelson arms, the length of the power recycling cavity, and the common length variations of the Fabry-Perot cavities. 
The relevant error signals also contribute to the longitudinal control signals sent to the different mirrors.

\begin{figure}[tb!]
\begin{center}
	\includegraphics*[viewport=0 0 447 150,width=0.8\linewidth]{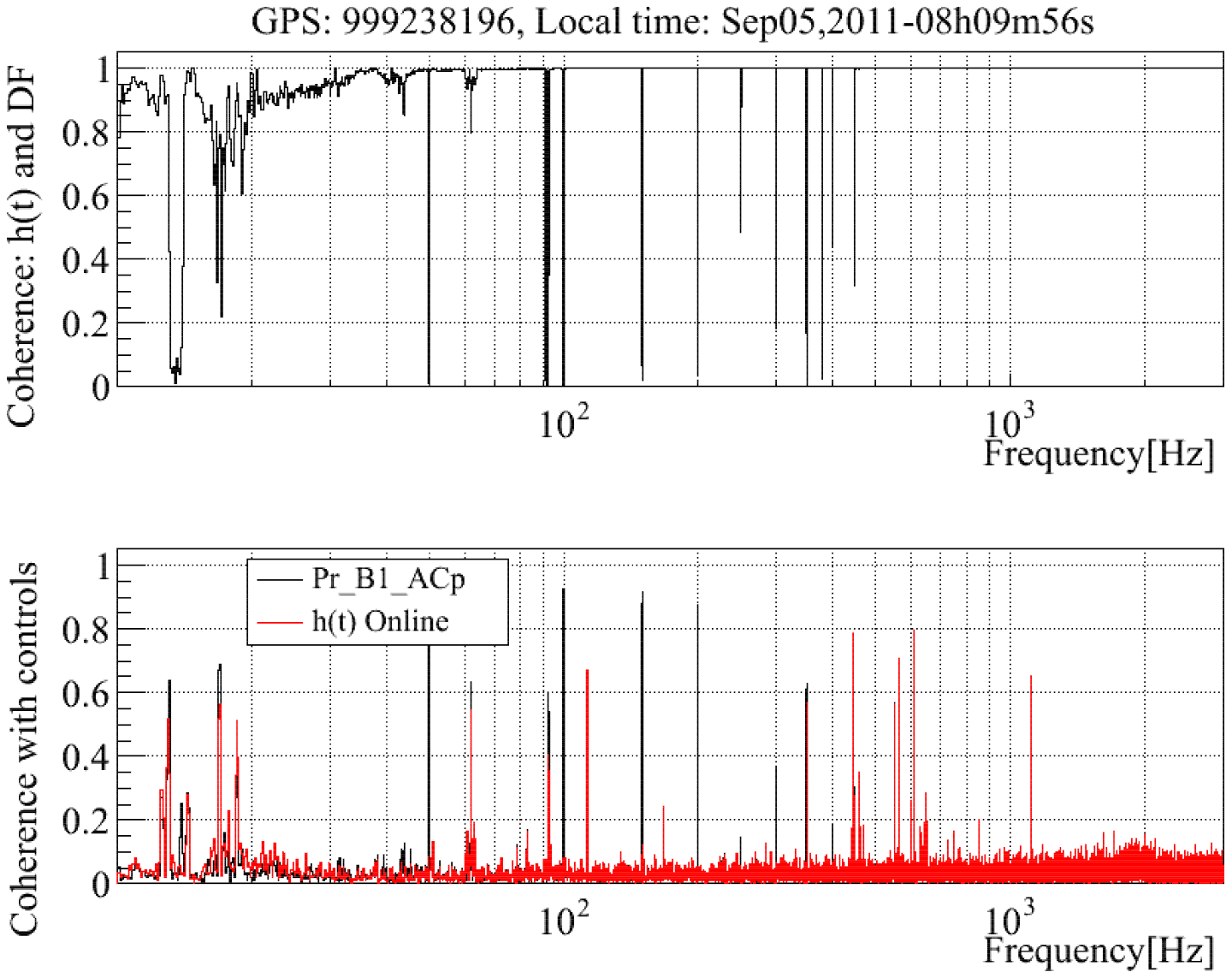} 
	\caption{
          Coherence between the sum of the control signals and $h(t)$ (red) or $\mathcal{P}_{AC}$ (black).
        }
	\label{fig:Cohe_hrec_vs_dof}
\end{center}
\end{figure}

If the control signals are not properly subtracted in the reconstruction process,
some residual coherence is expected between the $h(t)$ time-series and the measured auxiliary degrees of freedom of the ITF.
The sum of the coherences between the $h(t)$ channel and the three main auxiliary degrees of freedom
is shown in figure~\ref{fig:Cohe_hrec_vs_dof} (bottom). Except for the power lines, the coherence is pretty low,
indicating that the remaining control noise is small.
The behavior of $h(t)$ and of $\mathcal{P}_{AC}$ is about the same: 
it indicates that the control noises are already properly subtracted in the online loops
and that the reconstruction does not add extra-noise.

\subsubsection{Calibration line cancellation --}
\begin{figure}[tb!]
\begin{center}
	\includegraphics[width=0.8\linewidth]{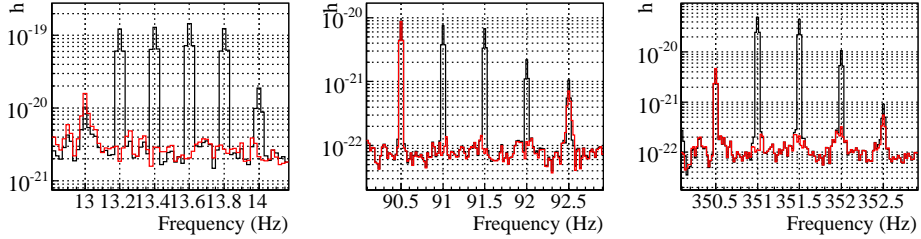} 
	\caption{Spectrum of $h(t)$ (red bold curve) and normalized spectrum of the dark fringe signal $\mathcal{P}_{AC}$ (black thin curve)
          around the three sets of calibration lines, with FFTs of 50~s.
          The frequencies of the calibration lines are summarized in the table~\ref{tab:CalibLines}.
          The lines at 90.5~Hz and 350.5~Hz are calibration lines from the photon calibrator, not subtracted in the $h(t)$ channel.
        }
	\label{fig:CalibLineCancellation}
\end{center}
\end{figure}

Another way to check that the reconstruction is working properly and that the control signals are properly subtracted
is to look at the residual amplitude of the calibration lines in $h(t)$.
The spectrum of $\mathcal{P}_{AC}$ and $h(t)$ around the three sets of calibration lines are shown in figure~\ref{fig:CalibLineCancellation}.

The optical gains and cavity finesse have been extracted from the set around $350\,\mathrm{Hz}$.
Therefore, a good cancellation is expected in this band, except if there is some phase error (time mismatch)
in the actuation or sensing models.
The cancellation is indeed compatible with the statistical limitations due to the finite signal-to-noise ratio of the calibration lines:
the NE and WE lines are cancelled at the 99\% level and the BS line at the 97\% level.
The cancellation factors at the other calibration lines are also compatible with statistics:
    97\% and better than 95\% for the NE and WE lines around $90\,\mathrm{Hz}$ and  $12\,\mathrm{Hz}$ respectively,
and 90\% and better than 75\% for the corresponding BS lines.
It indicates that the models are correct in the most critical frequency band of the reconstruction, where
the control signals and the dark fringe signals have similar contributions to $h(t)$. 

The PR control signals are cancelled by less than $\sim50\%$, due to their difference in model, 
but their contribution is much lower. As a consequence, they do not add a large fraction of extra-noise in $h(t)$.

\section{$h(t)$ uncertainty estimation}\label{lab:Errors}
The consistency checks described in the previous section have shown 
that no significant bias was found below 1~kHz in the amplitude and phase of the reconstructed $h(t)$ channel (section~\ref{lab:HrectoHinj})
and that the $h(t)$ time series does not contain extra-noise (section~\ref{lab:NoiseLevelInH}).

Below a few hundreds of hertz, the $h(t)$ channel is reconstructed as a complex combination of different and correlated signals 
after the application of different calibrated transfer functions.
It is thus difficult to estimate an uncertainty from the propagation of the individual uncertainties on the channels and their calibration.
A global way to estimate the uncertainty relies on the comparison of the reconstructed $h(t)$ signal with a calibrated signal $h_{inj}(t)$
injected into the detector as shown in section~\ref{lab:HrectoHinj}. This method only applies up to 1~kHz since the injected
signal is not calibrated at higher frequencies.
At higher frequency, the control signals contribute by less than 1\% to the $h(t)$ signal.
Therefore, in this frequency band, the systematic uncertainty comes only from the sensing model,
the uncertainty on the optical gain, and the uncertainty on the optical model which is small since we are well above the cavity pole.

The estimation of the systematic uncertainties on the amplitude and phase of the $h(t)$ time series in both frequency ranges are given below.

\subsection{Amplitude uncertainties}
Below 1~kHz, the comparison of $h(t)$ with $h_{inj}(t)$ shown in figure~\ref{fig:Hrec_to_Hinj_VSR4} is within 2\% in amplitude. 
The systematic uncertainty of the actuation model used to determine $h_{inj}$ is 5\%
and the error due to the long-wavelength regime approximation and the simple pole approximation of the optical response 
is lower than 0.5\%.
Therefore the systematic uncertainty on the $h(t)$ amplitude is 7.5\% below 1~kHz.

Above 1~kHz, the systematic uncertainty comes from:
\begin{itemize}
\item the optical gain, with an uncertainty of 6\%: 
statistical uncertainties of 1\% are estimated from the signal-to-noise ratio of the calibration lines used to extract the optical gains.
Moreover, the calibration uncertainty on the mirror actuators is of 5\%.
\item the sensing of the $\mathcal{P}_{AC}$ channel, with an uncertainty of 0.5\% in amplitude: 
the electronic response is flat to within better than 0.5\% in the 1~Hz--10~kHz band since the analog anti-aliasing filter
has a much larger cut-off frequency of around 100~kHz.
\item the shape of the optical response $O_{ITF}$: 1\%,
coming from the 6\% systematic uncertainties on the cavity finesse shown in section~\ref{lab:Check_Finesse},
\item the long-wavelength regime and simple pole approximation: 1\%
\end{itemize}
The sum of all the uncertainties gives an uncertainty of 8.5\% on the amplitude of the $h(t)$ time series 
above 1~kHz, slightly larger than at lower frequencies.

\subsection{Phase uncertainties}
As was the case for the amplitude uncertainty, the measurements shown in figure~\ref{fig:Hrec_to_Hinj_VSR4} indicate that the phase of $h(t)$
is properly reconstructed within 30~mrad below 1~kHz. 
The systematic uncertainty of the actuation model is 20~mrad.
Therefore, the systematic uncertainty on the $h(t)$ phase is 50~mrad below 1~kHz.

Above 1~kHz, the main systematic uncertainty comes from the timing calibration of the sensing of $\mathcal{P}_{AC}$,
estimated to be $4\,\mathrm{\mu s}$. 
The 6\% uncertainties on the cavity finesse induce less than $1.5\,\mathrm{\mu s}$ uncertainty in the $h(t)$ channel above 1~kHz.
As explained in section~\ref{lab:Check_Finesse}, the channel $\mathcal{P}_{AC}$ was delayed by $3\,\mathrm{\mu s}$
in order to match the correct finesse. This systematic bias must be added to the uncertainty on $h(t)$ timing.
As a consequence, the timing uncertainty on the $h(t)$ time series is estimated to be $\Delta t_d = 8\,\mathrm{\mu s}$.

Additionnally, due to the long-wavelength regime approximation and the simple pole approximation of the optical response,
the reconstructed $h(t)$ might be biased, by less than $4\,\mathrm{\mu s}$, depending on the sky direction of the GW.

\subsection{Uncertainty summary}\label{lab:UncertaintySummary}
The $h(t)$ reconstructed time series is valid from 10~Hz up to the Nyquist frequency of the channel used, 
i.e. up to 2048~Hz, 8192~Hz or 10000~Hz. 
In this validity range, the systematic uncertainty on the $h(t)$ amplitude is:
$$
\Delta A(f)/A(f) = 7.5\% \mathrm{\quad below\ 1\,kHz\quad and \quad }\Delta A(f)/A(f) = 8.5\% \mathrm{\ above} 
$$
Adding the timing systematic uncertainties and the low frequency phase uncertainty together, 
the systematic uncertainties on the phase of $h(t)$ can be defined, as a function of frequency: 
$$
\Delta \Phi(f) = (50\times10^{-3} + 2\pi f \Delta t_d)\ \mathrm{rad,}\quad\mathrm{with\ }\Delta t_d\,=\, 8\,\mathrm{\mu s}
$$
with an additional bias, lower than $\pm4\,\mathrm{\mu s}$, depending on the sky direction.

\section{Consistency checks with the photon calibrator}
In section~\ref{lab:HrectoHinj}, we showed how the reconstructed $h(t)$ can be checked with respect to
a differential arm length signal injected into the ITF. However, the check with signals injected through the electro-magnetic
actuators is limited, in particular a common error on the calibrated gains in all the actuators would not be detected
by this method.

The same principle can be used, but with an independent mirror actuator: the so-called photon calibrator (PCal).
Similar setups were also installed in GEO~\cite{bib:PCal_GEO} and LIGO~\cite{bib:PCal_LIGO,bib:PCal_LIGO_2,bib:PCal_LIGO_3}.
The principle of the PCal setup is described in this section and a simple consistency check of $h(t)$ performed
up to $\sim 400\,\mathrm{Hz}$ is done. 
Then, a more complicated analysis used to check $h(t)$ up to $\sim 6\,\mathrm{kHz}$ is detailed.

\subsection{Principle and calibration of the photon calibrator}
The PCal is based on a $\sim 1\,\mathrm{W}$ auxiliary laser (of wavelength $915\,\mathrm{nm}$) 
used to apply a force on a mirror by radiation pressure.
In Virgo, the setup was installed around the NI mirror such that the PCal beam pushes NI 
from the inner side of the Fabry-Perot cavity.
A photodiode is used to monitor the power of the auxiliary laser reflected by the NI mirror, $P_{ref}$, 
and thus to estimate the force $F$ applied on the mirror as:
\begin{eqnarray}
F &=& \frac{2\cos(i)}{\mathrm{c}}\,P_{ref}
\end{eqnarray}
where $i$ is the incidence angle of the auxiliary laser on the mirror and $\mathrm{c}$ the speed of light.\\

The setup is such that the auxiliary laser beam is generated outside the vacuum chamber 
and sent towards the center of the NI mirror through a viewport. 
The NI mirror reflects $\sim90\%$ of the incident beam power,
and $\sim10\%$ of the beam is transmitted. 
The reflected and transmitted beams are measured outside the vacuum chamber through two other viewports. 
The transmission coefficients of the viewports and their homogeneity were measured with the auxiliary laser before 
their installation on the vacuum chamber.

The monitoring photodiode was calibrated as a function of the reflected power $P_{ref}$,
using a NIST-calibrated power-meter.
Systematic uncertainties of $\sim5\%$ on $P_{ref}$ has been estimated from the power losses
measured between the injected power and the transmitted and reflected powers.
Different measurements for the power calibration of the photodiode were done in October 2010, after VSR3, and in November 2011, after VSR4.
Each calibration campaign resulted in a systematic uncertainty on $P_{ref}$ close to $5\%$,
and the calibration was found to be stable to within $2\%$ during one year.

The angle of incidence was estimated at $37.48^{\circ}\pm0.41^{\circ}$ during VSR3 and VSR4,
leading to $0.6\%$ uncertainty on the force. This study has also shown that the PCal beam hits
the NI mirror center to within 2~cm.

As a consequence, the systematic uncertainty on the force applied on the NI mirror estimated from the time-series 
monitoring $P_{ref}$ is between 5\% and 6\%.

Additionally, the delay introduced by the photodiode readout electronics has been measured
to be $51.2\pm1.0\pm4.0\,\mathrm{\mu s}$, where the $4.0\,\mathrm{\mu s}$ uncertainty coming
from the calibration of the Virgo timing system~\cite{bib:VirgoCalib} is the same for all the time-series of the Virgo data.\\

The force is applied modulating the PCal laser power between a few tens of hertz and a few~kHz. 
The NI mirror motion $\Delta x$ induced by a sinusoidal force of amplitude $\Delta F$ at frequency $f$ 
applied on the mirror can then be estimated from the
mechanical model $\mathcal{H}$ of the suspended mirror of mass $m=21.32\pm0.02\,\mathrm{kg}$    
with cables of length $l=0.7\,\mathrm{m}$: $\Delta x = \mathcal{H}\times \Delta F$.
Assuming that the mirror is a rigid body, which is valid for frequencies below $\sim 400\,\mathrm{Hz}$,
the mirror motion is, at frequencies well above the pendulum resonance frequency $f_0=0.6\,\mathrm{Hz}$,
\begin{eqnarray}
 \Delta x_{rigid}(t) &=&  -\frac{1}{m}\frac{1}{(2\pi f)^2} \times \Delta F(t)\ =\ -\frac{1}{m}\frac{2\cos(i)}{\mathrm{c}}\frac{\Delta P_{ref}(t)}{(2\pi f)^2}
\end{eqnarray}
where $\Delta P_{ref}(t)$ is the calibrated amplitude of the power reflected by the mirror
and $f$ is the PCal laser modulation frequency. The limitation of this model was first seen in the GEO interferometer~\cite{bib:PaperPCalGEO}.\\

Finally, the optical response of the ITF to a motion of the NI mirror must be taken into account in order to estimate the
equivalent strain $h_{inj}(t)$ injected into the ITF via the PCal.
The motion of the NI mirror modifies both the differential arm length, as expected, but also the length
of the short Michelson. 
As a consequence, the optical gain of the NI mirror motion is slightly lower than for the end mirrors.

\subsection{Validation of the sign of $h(t)$}
The sign of the $h(t)$ channel must be consistent between the different detectors 
since coherent analysis of their data are sometimes performed when searching for gravitational wave sources.
The $h(t)$ channel was defined in common with LIGO as:
\begin{eqnarray}
h(t) &=& \frac{L_x-L_y}{L_0}
\end{eqnarray}
where $L_x$ and $L_y$ are respectively the north and west arm lengths for Virgo.

The PCal setup was installed around the NI mirror such that the force pushes NI towards the exterior of the Fabry-Perot cavity.
Above its resonance frequency of $0.6\,\mathrm{Hz}$, the pendulum mechanical response induces a phase shift of $-\pi$
between the force applied on the mirror and the mirror motion.
Therefore, when the force increases towards the exterior of the cavity, the NI mirror moves such that the
cavity length $L_x$ decreases.
It has been checked that the phase of the transfer function from the force to $h(t)$ is $-\pi$
within a statistical uncertainty of $\sim1\,\mathrm{mrad}$. 
It confirms that the sign of the reconstructed $h(t)$ time-series is correct.
Note that this result does not depend on the calibration of the PCal.

\subsection{Simple consistency check below 400~Hz} \label{lab:PCal_Check1}
\begin{figure}[tb!]
\begin{center}
	\includegraphics[width=0.8\linewidth]{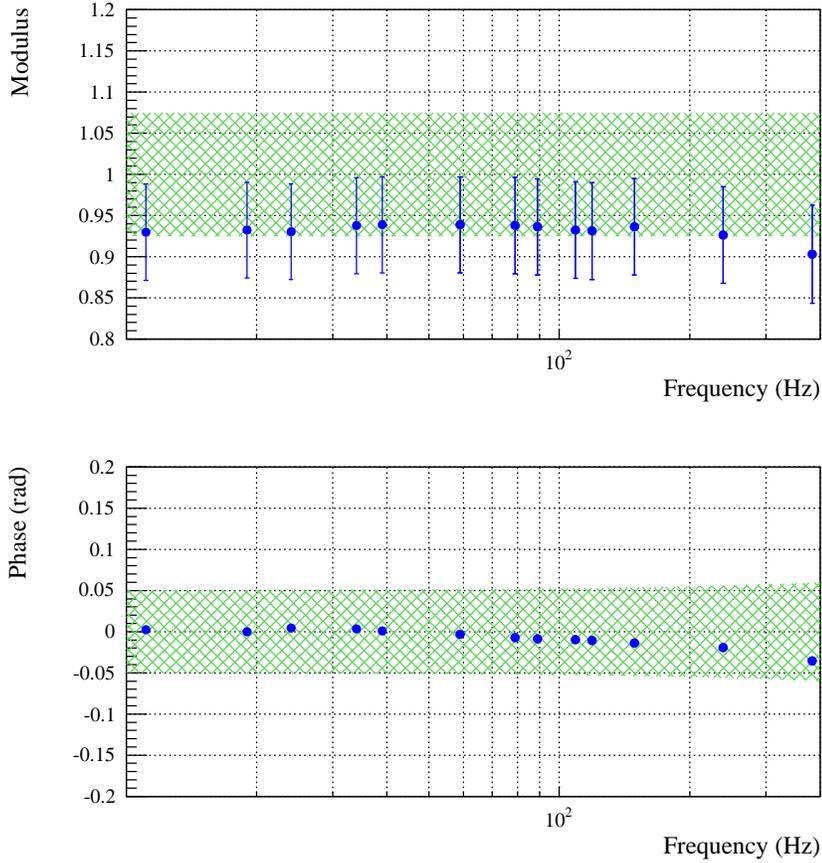}
	\caption{Comparison of $h_{rec}$ to $h_{inj}$ during VSR4.
          The statistical errors are negligible.
          Error bars indicate the systematic uncertainty of $h_{inj}$, 
          assuming that the mirror is a rigid body.
          The colored areas indicate the systematic uncertainties of $h_{rec}$
          (without the $4\,\mathrm{\mu s}$ that are present in both signals).
        }
	\label{fig:PCal_Check1}
\end{center}
\end{figure}

Injections of signals between 10~Hz and 1~kHz in the ITF via the PCal were carried out 
during a campaign after VSR3, and every week during VSR4.
The comparison of the reconstructed channel $h_{rec}$ to the injected strain $h_{inj}$
is shown in figure~\ref{fig:PCal_Check1}, 
using the simple mechanical model described above.
This model allows to convert the force applied with the PCal on the NI mirror
to an equivalent strain signal $h_{inj}$ without any free parameter,
up to $\sim 400\,\mathrm{Hz}$.
The amplitude ratio is found to be~0.92 during VSR3 and~0.93 during VSR4,
with uncertainties of the order of $0.10$, 
and the phase difference is close to~0, with uncertainties of the order of 50~mrad.

In conclusion, the analysis of the PCal injections has validated the
$h(t)$ reconstruction and the associated uncertainties in the range 10~Hz to 400~Hz.

\subsection{Large-band consistency check, up to 6~kHz} \label{lab:PCal_Check2}
PCal injections were performed up to 1~kHz during VSR3, and up to 6~kHz during VSR4 and the fall of 2011.
In order to analyze these injections, a more complete model of the mechanical response of
the suspended mirror must be used: the internal deformation modes of the mirror have to be
taken into account~\cite{bib:PaperPCalGEO}.

\subsubsection{The mechanical model -- }
In this model, the effective longitudinal motion of the mirror center is described as the sum
of its motion as a rigid body and of the effective motions due to the internal modes $m$:
\begin{eqnarray}
\Delta x_{eff} &=& \Delta x_{rigid}\ + \ \sum_{m} \Delta x_m 
\end{eqnarray}
For a given mode, the effective motion induced by a sinusoidal force of amplitude $\Delta F$ at frequency $f$ 
and pushing the mirror at position $\vec{r}$ can be described as:
\begin{eqnarray}
\Delta x_m(f) &=& \mathcal{R}_m(I,\mathcal{G}) \times \mathcal{G}_m \times \mathcal{C}_m(\vec{r},f) \times \Delta F(\vec{r},f)
\end{eqnarray}
where:
\begin{itemize}
\item $ \mathcal{G}_m$ is the ``geometry'' of the mirror surface deformation for the mode.
The main modes of the mirror~\cite{bib:NoteMirrorModes} are
(i) the drum modes which have a maximum of the deformation amplitude at the center of the mirror,
and (ii) the butterfly modes which have a null deformation at the center of the mirror.
\item $\mathcal{C}_m(\vec{r},f)$ is the coupling between the applied force and the mirror mode at frequency $f$.
The frequency dependent part of the coupling is described as a second-order low-pass filter with the cut-off frequency $f_{0,m}$ at
the resonant frequency of the mode and a quality factor $Q_m$. 
The absolute coupling level depends on the distribution of the force on the mode geometry: as the PCal beam hits the
mirror at its center to within 2~cm, the drum modes must be excited while the butterfly modes excitation must be low.
The parameters of the mirror internal modes were estimated with finite-element simulations~\cite{bib:NoteMirrorModes}:
the first drum modes resonant frequencies $f_{0,m}$ are around $5670$~Hz and $15670$~Hz, with quality factors $Q_m$ 
of the order of $10^5$.
\item $\mathcal{R}_m(I,\mathcal{G})$ is the coupling between the mirror surface deformation and the interferometer.
The ITF beam illuminates only the central part of the mirror and senses its deformations  
with a weight function given by its Gaussian intensity distribution $I$.
As a consequence, the ITF converts the mirror deformation into an effective longitudinal mirror motion.
As $I$ is centered on the mirror within better than 1~mm, this coupling is strong for the drum modes but it is expected to be low
for the butterfly modes. 
Finite-element simulations have confirmed that they can be neglected.
\end{itemize}
The resonance of the higher order modes $n$, with high frequency, cannot be observed in the data sampled at 20~kHz,
but might contribute at low frequency through a frequency-independent gain between the force and the effective mirror motion.
This effective gain is given by the low-frequency part of $G_{HOM} = \sum_{n} \mathcal{R}_n \mathcal{G}_n \mathcal{C}_n$.

The violin modes of the suspended mirror are also part of such a model.
However, they contribute only close to their resonant frequencies and they are neglected in the following analysis.

\subsubsection{Determination of the mechanical model parameters --}

\begin{table}[htbp!]
\begin{center}
\caption{Results of the different fits of the mirror modes above 4800~Hz.}
\begin{tabular}{|l|l|c|c|c|c|}
\hline
Step                    & Mode   & $f_0$ (Hz)            & $Q$              & $G_{m}^{norm}$   &  $\mathcal{P}(\chi^2)$ \\
\hline
\multirow{3}[0]{*}{(i)} & Drum-  & $5671.2475\pm0.0001$  & $400300\pm6700$  & $0.1653\pm0.0004$              & \multirow{3}[0]{*}{35.1\%} \\
                        & Violin & $5674.2221\pm0.0024$  & $363000\pm43000$ & $0.0070\pm0.0003$              &  \\
                        & Drum+  & $5675.6054\pm0.0010$  & $188000\pm34000$ & $0.158\pm0.001$                & \\
\hline
(ii)                    & HOM    & --                    & --               & $0.670\pm0.014$              & 6.6\% \\
\hline
\end{tabular}
\label{tab:PCal_FitResults}
\end{center}
\end{table}

\begin{figure}[tb!]
\begin{center}
	\includegraphics[width=0.8\linewidth]{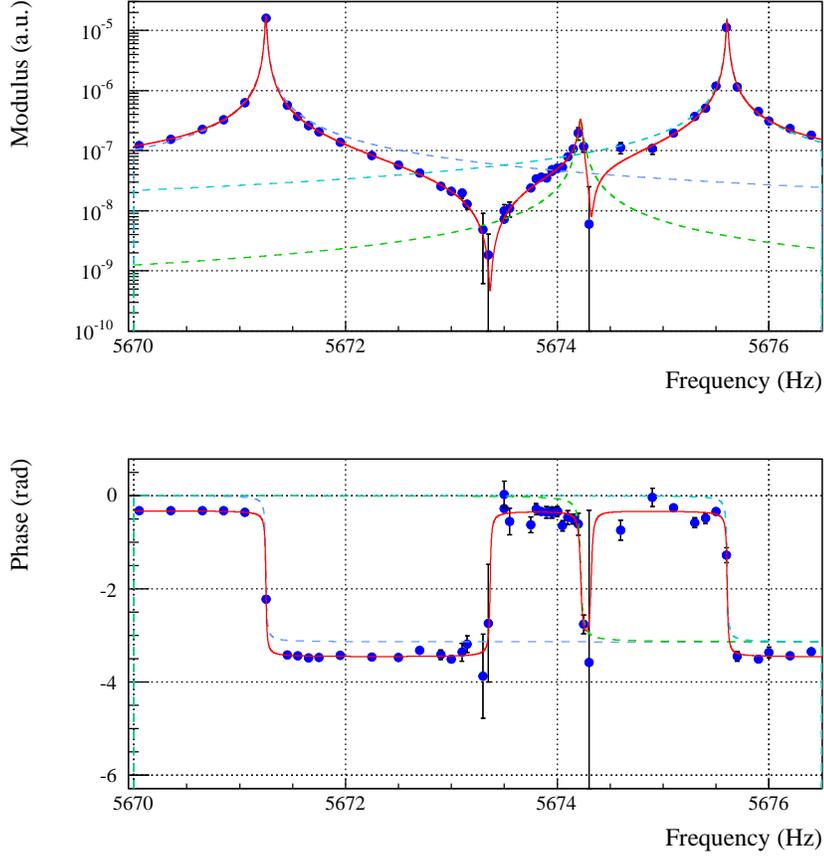}
	\caption{Transfer function $\mathcal{H}$ from the force applied on the mirror via the PCal
          to the mirror motion.
          The absolute values of the y-axis were not used in the analysis described in the paper.
          Blue points: data.
          Blue, light blue and green lines: fitted responses for the two drum modes and the violin mode of NI.
          Red line: full response.
        }
	\label{fig:PCal_DrumModes}
\end{center}
\end{figure}

At this point, the complete model that allows to convert the force applied with the PCal on the NI mirror
to an equivalent strain signal $h_{inj}$ has some unknown parameters: 
the resonant frequencies and quality factors of the mirror internal modes,
and their effective coupling ($\mathcal{R G C}$).
The PCal data confirmed that the first drum mode is visible. In practice,
as shown in figure~\ref{fig:PCal_DrumModes}, the drum mode
is split in two modes, separated by 5~Hz due to geometrical asymmetries of the mirror.
A third mode is visible in between: it has been identified as coming from a coupling
between the drum modes and the 13th violin mode. This violin mode had therefore been taken into account
in the analysis.

\begin{figure}[tb!]
\begin{center}
	\includegraphics[width=0.8\linewidth]{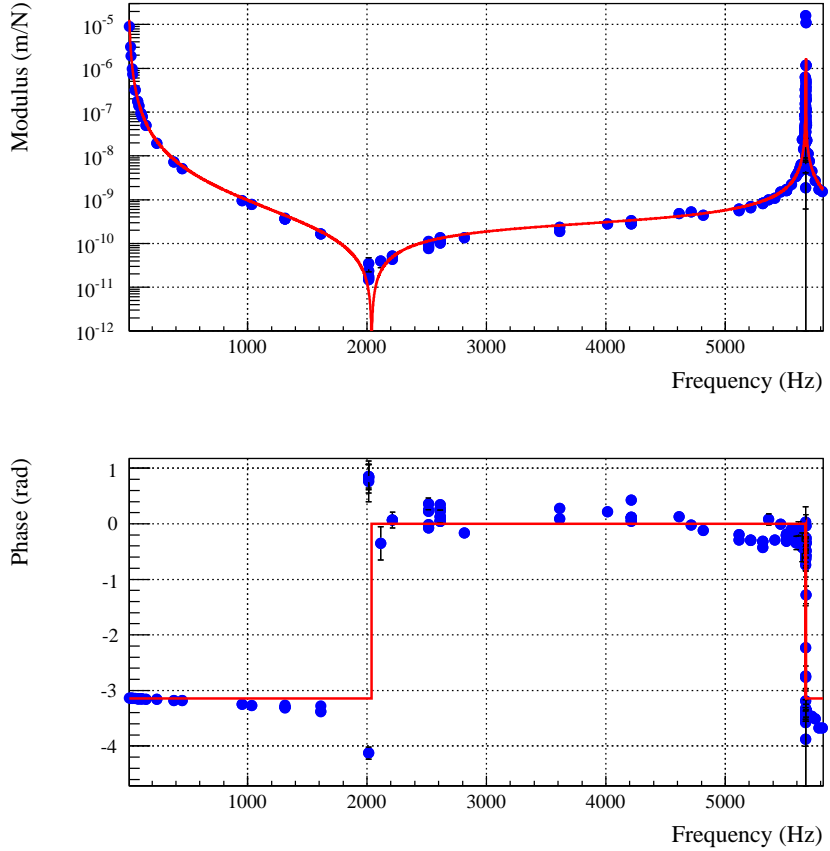}
	\caption{Mechanical model (red curve) for the PCal injections from 10~Hz to 6~kHz
          as measured by the ITF.
          The blue points are extracted from the PCal injections for comparison.
          Only the points between 4800~Hz and 5680~Hz and the notch frequency $f_N$ 
          were used to determine the free parameters of the model.
        }
	\label{fig:PCal_FullMechanicalModel}
\end{center}
\end{figure}

In this analysis, the only assumption concerning $h(t)$ is that any bias in the reconstructed channel is frequency-independent in the range
4800~Hz to 5680~Hz. This is reasonable since the ITF is free in this frequency range
such that the shape of the conversion from $\mathcal{P}_{AC}$ to $h(t)$ depends only on the dark fringe sensing $S$,
with a flat modulus within 0.5\% in this range, 
and on the ITF optical model $O_{ITF}$, with a simple $f^{-1}$ modulus shape in this range.
The analysis is done by adjusting the following free parameters around the resonance peaks seen in the transfer function $\mathcal{H}$
from the force applied on the NI mirror, estimated from the monitoring photodiode channel,
to the mirror motion, estimated from the $h(t)$ channel during VSR4:
\begin{enumerate}
\item the three resonant frequencies $f_{0,m}$ and three quality factors $Q_m$ of the three modes observed in the data,
as well as their three relative coupling factors $G_{m}^{norm}$, are fitted in the range 5670~Hz to 5677~Hz, 
on the transfer function\footnote{
A phase offset of -322\,mrad was added to match the phase in this range, which would corresponds to a delay of $9\,\mathrm{\mu s}$.} 
$\mathcal{H}$.
\item the coupling factor of the higher-order mirror modes relative to the three observed modes, $G_{HOM}^{norm}$
is fitted on the transfer function $\mathcal{H}$ in the range 4800~Hz to 5500~Hz. 
The resulting fits of steps (i) and (ii), as well as their $\chi^2$ probability, are summarized in table~\ref{tab:PCal_FitResults}.
\item at this point, the model can be written 
\begin{eqnarray}
\Delta x_{eff} &=& \Delta x_{rigid} + \\
               && G_{modes} \times \big( \Delta x_{Drum-}^{norm} + \Delta x_{Drum-Violin}^{norm} + \Delta x_{Drum+}^{norm} + \Delta x_{HOM}^{norm} \big) \quad \nonumber
\end{eqnarray}
where the only remaining free parameter is $G_{modes}=\sum_{m} G_{m}$. 
It can be determined precisely from the frequency of the notch observed at $f_N = 2035\pm2$~Hz since $\Delta x_{eff}(f_N) = 0$.
Note that this determination is completely independent of the $h(t)$ reconstruction as it can be estimated
from the $\mathcal{P}_{AC}$ channel.
It results in $G_{modes} = (2.714\pm0.006)\times10^{-10}\,\mathrm{m/N}$.
\end{enumerate}
After this analysis, the mechanical model of the PCal response is determined without any free parameters in the range 10~Hz to 6~kHz.
It is shown and compared to the data in the figure~\ref{fig:PCal_FullMechanicalModel}.
One can thus estimate the equivalent strain $h_{inj}=\frac{\Delta x_{eff}}{L}$ injected via the PCal up to 6~kHz.

\subsubsection{Analysis of the PCal injections --}
\begin{figure}[tb!]
\begin{center}
	\includegraphics[width=0.8\linewidth]{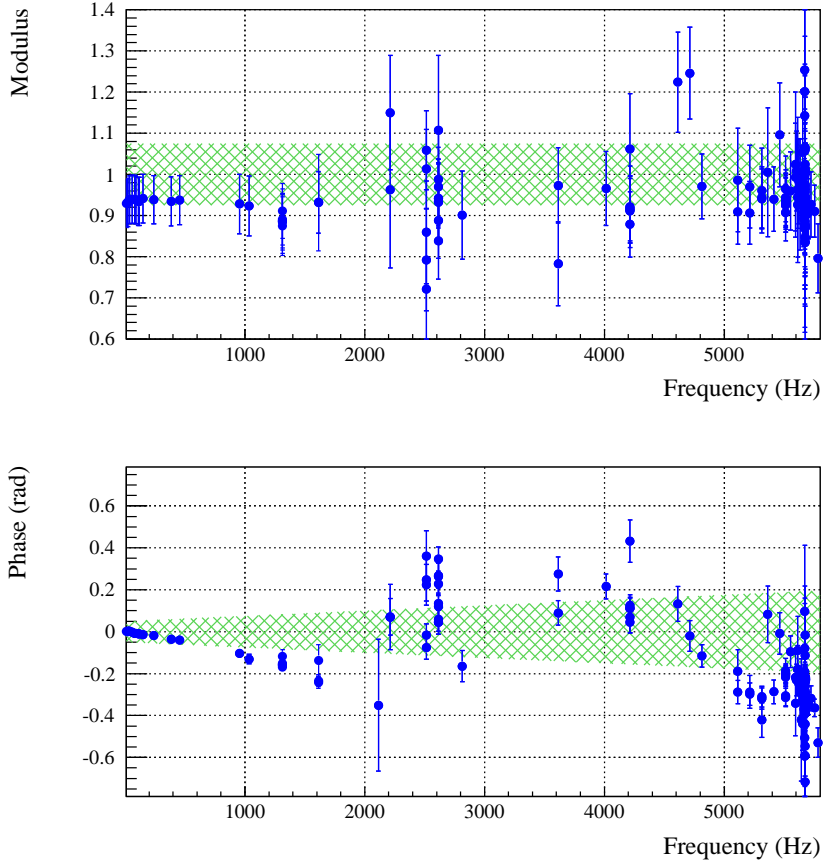}
	\caption{Comparison of $h_{rec}$ to $h_{inj}$ during VSR4.
          $h_{inj}$ was estimated using the full mechanical model that includes the mirror internal modes.
          Error bars indicate the systematic uncertainty of $h_{inj}$.
          The colored areas indicate the systematic uncertainties of $h_{rec}$ 
          (without the $4\,\mathrm{\mu s}$ that are present in both signals). 
        }
	\label{fig:PCal_Check2}
\end{center}
\end{figure}

The comparison of the reconstructed channel $h_{rec}$ to the injected strain $h_{inj}$ during VSR4,
using the mechanical model extracted in the previous section,
is shown in figure~\ref{fig:PCal_Check2}.
The amplitude ratio is found to be flat around~0.93 with uncertainties of the order of~0.10.
The phase difference is close to~0, with uncertainties of 
the order of 50~mrad at low frequency.
At few kilohertz, the bias due to the use of the simple pole approximation for the response to a mirror motion 
is expected to start being visible in this comparison as a delay of less than $10\,\mathrm{\mu s}$.
A small discrepancy, of the order of $2\,\mathrm{\mu s}$, between $h_{rec}$ and $h_{inj}$ phases appears above 5~kHz, 
but the fluctuations of the measured phase do not allow to firmly conclude whether it is linked to this bias.

In conclusion, the independent check of $h(t)$ with the PCal was performed up to 6~kHz during VSR4. 
It has shown that the amplitude and phase of $h(t)$ are correct, as well as the associated uncertainties.

\subsection{Stability of $h(t)$ during VSR4}
Two calibration lines were sent to the ITF through the NI PCal during VSR4,
at 90.5~Hz and 350.5~Hz, with a signal-to-noise ratio of the order of 100 and 60 respectively
when integrated over 10~s 
(in addition to the lines described in table~\ref{tab:CalibLines}).
Such permanent injections allow to study the stability of both the reconstruction of $h(t)$
and the PCal setup and calibration.
In particular, during VSR4, two hardware modifications could have had impact on the reconstruction:
(i) the actuation electronics~$A_i$ was modified on July 19th 2011,
and (ii) the dark fringe photodiode readout electronics~$S$ was modified to be able to switch 
between the initial electronics path and a path with an additional filter. The switch between the
two configurations was automated, starting from June 16th 2011. 

The transfer function from the PCal monitoring photodiode channel to the $h(t)$ channel
was measured at both frequencies and averaged over 2~minutes. The data with a coherence
between both channels higher than 0.9999 were selected.
The modulus and phase of the transfer function at both frequencies were stable during VSR4
within statistical uncertainties of $1\%$ in modulus and $9\,\mathrm{\mu s}$ in phase.
These measurements imply that the reconstruction was stable to within these statistical uncertainties
during VSR4, including the $h(t)$ reconstruction updates following the hardware modifications.

\section{Summary}
The Virgo method used to reconstruct the gravitational wave strain $h(t)$ from the interferometer data
has been described. One of its main features is that it does not rely on the evolution of the filters
of the global control of the interferometer. Another advantage is that the control noises
are subtracted, with rejection factors of the order of~100.
The reconstruction was running during the four science data taking periods between 2007 and 2011,
with a dead-time as low as 0.07\%.
The reconstructed $h(t)$ channel and its associated amplitude and phase uncertainties have been validated 
with several consistency checks, as well as with the independent calibration method 
based on the photon calibrator installed in Virgo. From 10~Hz to 1~kHz, 
the systematic uncertainties have been estimated at 7.5\% in amplitude, increasing up to 8.5\% at 10~kHz.
The phase systematic uncertainties have some frequency dependence, 
starting from 50\,mrad at 10~Hz and increasing following a delay of $8\,\mathrm{\mu s}$ at high frequency.
An additionnal bias, lower than $4\,\mathrm{\mu s}$, depending on the sky direction of the GW, is present
due to the combination of the long-wavelength interferometer and the simple pole cavity approximations.
This is well within the constraints given by the data analysis on the calibration and reconstruction uncertainties.
Moreover, the $h(t)$ channel was available online for data analysis pipelines with a latency of 20~s.

The calibration uncertainties of the Virgo photon calibrator are slightly larger than that of the standard calibration methods.
It allows a direct check of the sign of the reconstructed $h(t)$ channel compared 
to the definition taken in agreement with other experiments such as LIGO
and a simple validation of the reconstruction and its uncertainties up to a few hundreds of hertz.
Moreover a more detailed analysis of the Virgo PCal data allowed to validate $h(t)$ above 1~kHz,
which cannot be done by the other techniques. 
The injection of permanent calibration lines in the interferometer with
the PCal also permit to confirm the robustness of the calibration and $h(t)$ reconstruction with respect to different
hardware modifications during VSR4.

The same reconstruction method is intended to be used for the next generation of detector currently under construction, Advanced Virgo.
If data analysis requires the latency be reduced below a few tens of seconds, 
the method will still be applied but in the time-domain directly.
The upgrade of the method still has to be studied to adapt it to the addition 
of the signal-recycling mirror in the second phase of Advanced Virgo.\\
The independent calibration method using the PCal will also be upgraded for Advanced Virgo, 
with the goal of reducing the power calibration uncertainties.

\appendix

\section{Control noise subtraction in the $h(t)$ reconstruction}\label{lab:CtrlNoiseSubtraction}

In total, four longitudinal degrees of freedom are controlled in Virgo,
defined from the lengths shown in figure~\ref{fig:OpticalScheme}:
\begin{itemize}
\item $\Delta L$ (or Darm), differential arm length, $L_N - L_W$,
\item Carm, common arm length, $L_N + L_W$,
\item Mich, differential length of the short Michelson: $l_N - l_W$,
\item Prcl, length of the power-recycling cavity: $l_{PR} + \frac{l_N + l_W}{2}$
\end{itemize}

The $h(t)$ reconstruction combines different channels where the control noises might be present
in such a way that the noise is subtracted.
As stated in section~\ref{lab:LongLoop}, techniques of online control noise subtraction were successfully applied in Virgo.
As a consequence, as shown in section~\ref{lab:NoiseLevelInH}, the control noise was not limiting sensitivities, 
estimated in the frequency-domain directly, or estimated from $h(t)$.\\

\begin{figure}
\begin{center}
\label{fig:NoiseSub_Sensitivities}
  \subfigure[Time- and frequency-domain sensitivities without online noise subtraction.] {
    \label{fig:NoiseSub_Sensitivities_NoNoiseSubtraction}    
    \includegraphics[width=0.4\linewidth]{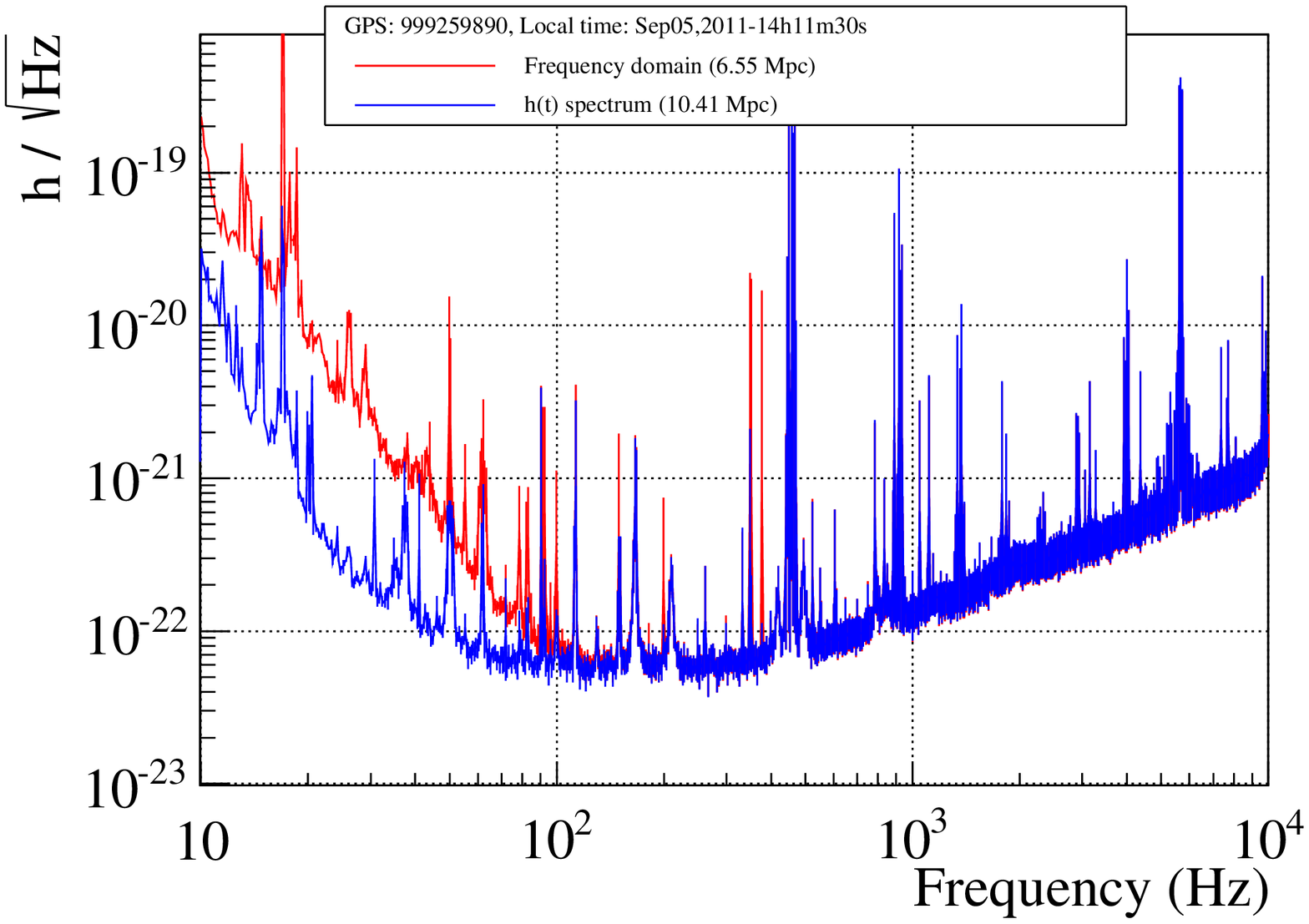} }
  \subfigure[Time-domain sensitivities with and without online noise subtraction.] {
    \label{fig:NoiseSub_Sensitivities_W_WO_Filters}
    \includegraphics[width=0.40\linewidth]{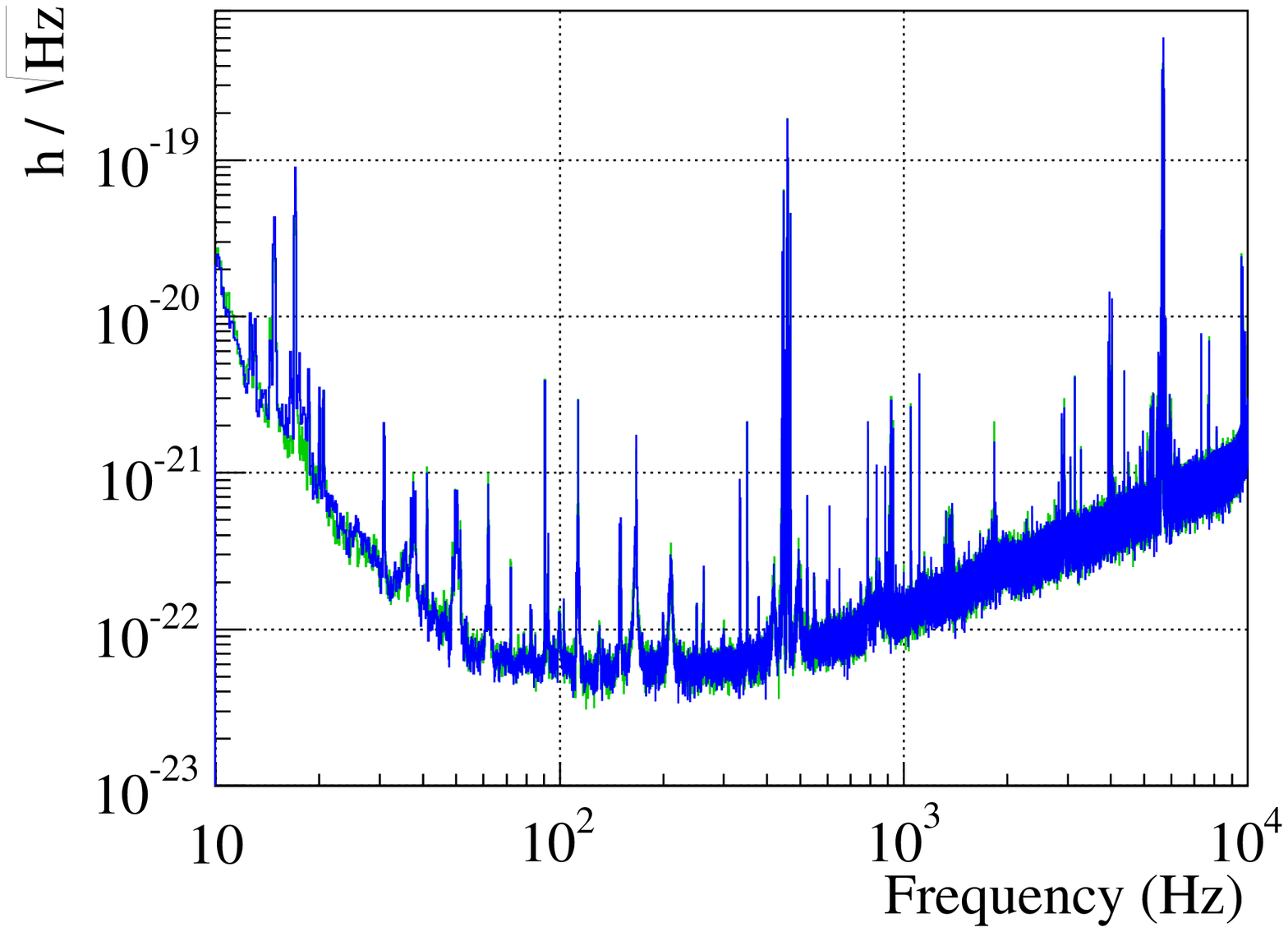} }
  \caption{(a) Sensitivity curves measured, at the same time, in the frequency-domain (red) and in the time-domain (blue)
    with the online noise subtraction disabled. 
    (b) Sensitivity curves measured in the time-domain
    with the online noise subtraction enabled (blue) and disabled (green),
    in two datasets separated by two~minutes.
  }
\end{center}
\end{figure}

In order to characterize further the performance of the reconstruction with respect to
control noise subtraction, specific data were taken right after VSR4 where the control noise cancellation
techniques were temporarily disabled.
As expected in this case, the sensitivity computed in frequency-domain 
was limited by control noise below $\sim100\,\mathrm{Hz}$ as shown in figure~\ref{fig:NoiseSub_Sensitivities_NoNoiseSubtraction}. 
On the contrary, as shown in figure~\ref{fig:NoiseSub_Sensitivities_W_WO_Filters}, the sensitivity computed from the time-domain channel $h(t)$
was still the same as when the control noise was subtracted online in the control loops.
This indicates that the control noise is properly subtracted in the reconstruction.
As a consequence, the ratio of the two estimated sensitivities gives an upper limit
of the precision of the control noise subtraction in the $h(t)$ reconstruction:
the residual contribution of the control noise in $h(t)$ is below a few percents.

\begin{figure}
\begin{center}
	\includegraphics[width=0.8\linewidth]{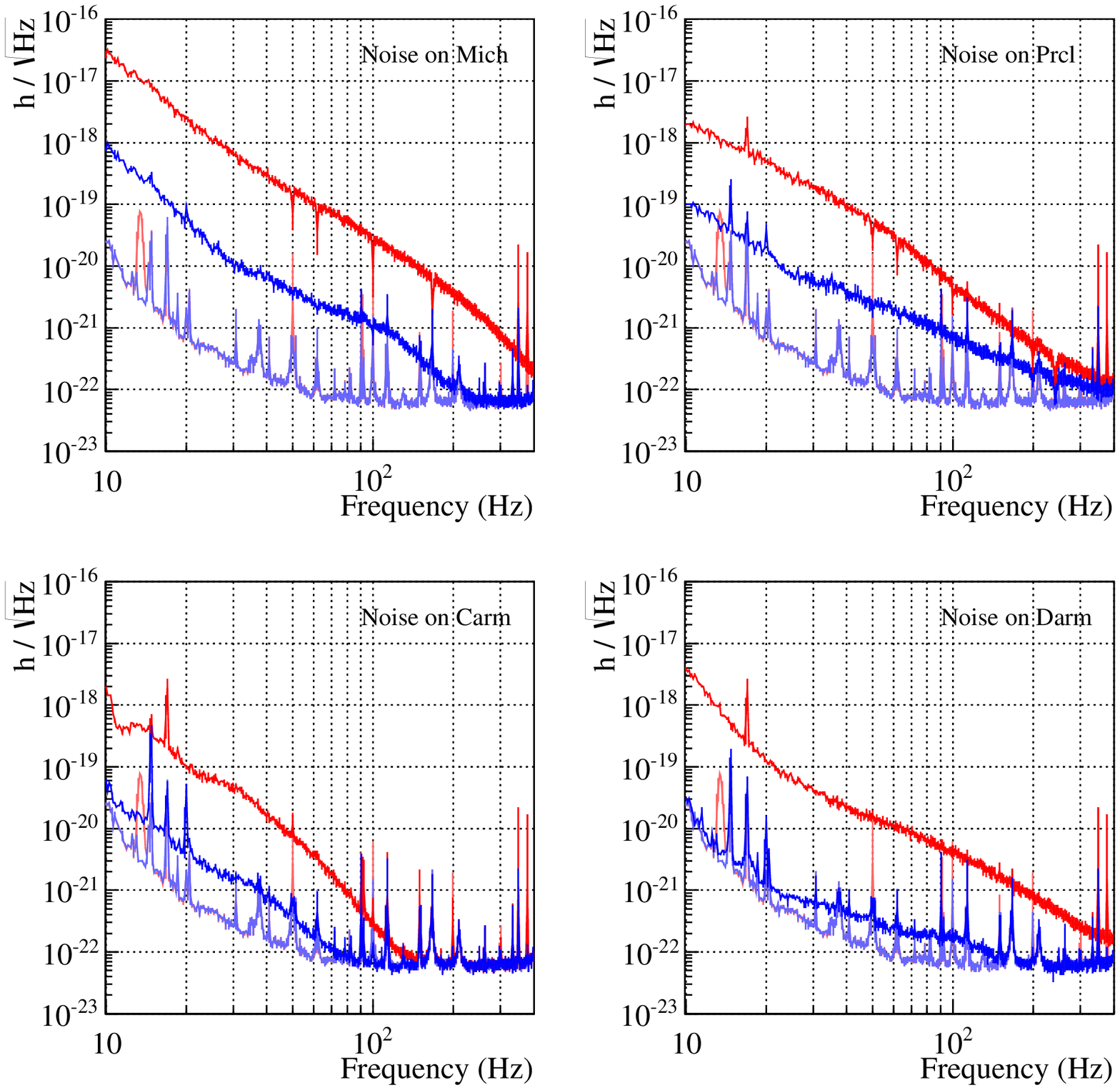} 
	\caption{Sensitivity curves estimated in the frequency-domain (red) 
          and in the time-domain (blue).
          Lighter lines indicate datasets with online noise cancellation 
          and no noise injected.
          Deeper lines indicate datasets without online noise cancellation 
          and with injection of control noises in the ITF.
        }
	\label{fig:NoiseSub_Sensitivities_withControlNoise}
\end{center}
\end{figure}

In order to better estimate the noise subtraction performances in the $h(t)$ reconstruction,
additional datasets were taken with the online noise subtraction technique disabled
and extra-noise added to the different controlled degrees of freedom (Mich, Prcl, Carm and $\Delta L$).
The measured sensitivity curves estimated from the frequency-domain and time-domain computations are
shown in figure~\ref{fig:NoiseSub_Sensitivities_withControlNoise}, as well as the reference
sensitivities measured with no noise injected.
In this case, extra-noise is visible in the reconstructed $h(t)$ channel, which allows
to estimate the precision of the subtraction.
Only a few percent of the control noise is present in the $h(t)$ channel up to at least $\sim100~\mathrm{Hz}$.
The $\Delta L$ degree of freedom is even rejected by more than a factor 100 below $\sim20\,\mathrm{Hz}$.
The subtraction of the noise in the Prcl degree of freedom is the least efficient,
as expected since it is known that the optical response model for PR mirror is approximate at low frequency.

Note that during these measurements, the signal-to-noise ratio of the calibration lines was lowered
by a factor~2 to~3 due to the injected noise. As a consequence, the noise subtraction might have been
slightly less efficient than in standard conditions.\\

To conclude, in this appendix, the $h(t)$ reconstruction was studied when using the actuation and sensing
parameterization given by the calibration procedures. It resulted in noise rejection factors of the order of~100.
Fine-tuning of the parameterizations could be done in order to further improve the noise subtraction
in the $h(t)$ signal reconstruction.

\end{document}